\documentclass[aps,preprintnumbers,superscriptaddress,nofootinbib,aps,prd,floatfix,twocolumn]{revtex4-2}
\pdfoutput=1 
\usepackage{graphicx}% Include figure files
\usepackage{amsthm} 
\usepackage{amsmath,amssymb,physics}
\usepackage{url} 
\usepackage{lineno} %Fred added
\usepackage{xspace}  %Fred added
\usepackage{braket}  %Fred added
\usepackage{hyperref}
\usepackage{xcolor} 
\usepackage{comment} %Max added
\usepackage{array} %Lucas added
\usepackage{%color,       
            soul}
\usepackage{longtable}
\usepackage{xparse}
\usepackage{subfigure}
\usepackage{widetable}
\usepackage{xparse}
\usepackage{subfigure}
\usepackage{stackengine}

\newcommand{\orcid}[1]{\,\href{https://orcid.org/#1}{\includegraphics[width=9pt]{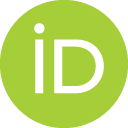}}}

\newcommand{\orcidLK}{0009-0007-5639-0350} % LUCAS

\definecolor{darkblue}{rgb}{0.0,0,0.5}
\definecolor{darkgreen}{rgb}{0.0,0.3,0.0}
\definecolor{redish}{rgb}{0.675,0,0.2}
\definecolor{red}{rgb}{0.8,0,0}
\definecolor{green}{rgb}{0,0.6,0}
\definecolor{blue}{rgb}{0,0,0.8}

\def\smu{{Department of Physics, Southern Methodist University,
    Dallas, TX 75275-0175, USA \looseness=-1}}

\NewExpandableDocumentCommand\mcc{O{1}m}
    {\multicolumn{#1}{l}{#2}}
\usepackage{multirow}
\newcolumntype{C}[1]{>{\centering\let\newline\\\arraybackslash\hspace{0pt}}m{#1}}

\def\smu{{Department of Physics, Southern Methodist University,
    Dallas, TX 75275-0175, USA \looseness=-1}}

\begin{document}

\title{A study of experimental sensitivities to proton parton distributions with xFitter}

\author{Lucas Kotz\orcid{\orcidLK}}\affiliation{\smu}

\date{\today}

\begin{abstract}
In collider physics, parton distribution functions (PDFs) play a crucial role in computing theoretical cross sections for scattering reactions. This study explores how different experimental data sets influence extracted PDFs in CTEQ-TEA and MSHT NNLO PDF analyses. To gauge the impact of experimental data, including the HERA and ZEUS combined charm and beauty production, LHCb 7 TeV charm and beauty production, CMS 7 TeV W+c production, and CMS 13 TeV inclusive jets, I utilize the $L_2$ sensitivity statistical indicator in the Hessian framework as a visual representation of their respective impacts. This sensitivity quantifies the statistical pulls on individual data sets against the best-fit PDFs, facilitating the identification of tensions among competing data sets. Using the QCD fitting framework xFitter, I extract the necessary values for plotting $L_2$ sensitivities for ten distinct data sets implemented in the program, employing recent PDF sets from the CTEQ-TEA and MSHT groups. The computed $L_2$ sensitivities estimate the potential impact of the examined data sets.
\end{abstract}

\maketitle

\section{Introduction}
\label{Introduction}

Parton distribution functions (PDFs) play a large part in collider physics as they are used to calculate hard cross sections in perturbative QCD \cite{RevModPhys.67.157,BlackBook,FoundationsofQCD}. Various groups study the impact of experimental data on a global QCD fit, which are performed to determine PDFs \cite{NNPDF:2014otw,Dulat:2015mca,Harland-Lang:2014zoa,NNPDF:2017mvq,Alekhin:2017olj,ATLAS:2012sjl,Bailey:2020ooq,H1:2015ubc,NNPDF:2021njg}. The choice of experimental data is crucial, as each experiment studies specific reactions that most directly assess the PDFs within a defined region of $x$ and $Q$, where $x$ is the momentum fraction of a given parton, and $Q$ is the factorization scale.

Recent fits, performed at the next-to-next-to-leading-order (NNLO) in the QCD coupling strength $\alpha_s$, can utilize either the Hessian \cite{HERAFitterdevelopersTeam:2014fzy,Bailey:2020ooq,Hou:2019gfw,Hou:2022onq,H1:2015ubc} or Monte Carlo (MC) \cite{NNPDF:2017mvq,NNPDF:2014otw,NNPDF:2021njg} framework to analyze the uncertainty on the PDFs. Determining which experiments to include in these fits demands a substantial effort. For each global analysis group, including and studying the impact of a new dataset in their fitting code requires a high level of expertise and a considerable amount of time. To hasten the investigation of the experimental impact on a specific PDF set, I use the $L_2$ sensitivity to visually display the potential influence of an experiment on a PDF set given $x$ and $Q$  instead of directly including the data set in the PDF set \cite{Hobbs:2019gob,Jing:2023isu}.

The $L_2$ sensitivity method simplifies our comprehension of statistical influences by directly mapping how each data set affects a given PDF. It assesses the correlation between each data set and the PDF, as well as the degree to which the data set impacts the determination of the PDFs. Plotting $L_2$ sensitivities on the $y$-axis, for a given $Q$, allows for the visualization of the pulls exerted by each experiment.

 I select PDF sets from the CTEQ-TEA and MSHT groups specifically NNLO CT18 \cite{Hou:2019gfw}, asymmetrical-strange CT18As \cite{Hou:2022onq}, and MSHT20 \cite{Bailey:2020ooq} sets. xFitter \cite{HERAFitterdevelopersTeam:2014fzy}, a publicly distributed framework for performing PDF fits, implements multiple data sets, encompassing a small subset of those already included in the PDF sets and those that are not yet. This enables evaluation of potential impacts arising from the multiple data sets included in xFitter, along with those not yet in the CT18 analysis, such as the H1+ZEUS combined $c$ and $b$ production \cite{H1:2018flt}, CMS W+c production \cite{CMS:2013wql}, and LHCb charm and bottom production \cite{LHCb:2013xam,LHCb:2013vjr}. The values necessary to calculate the $L_2$ sensitivity have been taken from the xFitter output.

\section{\(L_2\) sensitivity}
\label{L2_sensitivity}
The $L_2$ sensitivity method assesses how experiments impact a given PDF. It also visualizes statistical tensions among the experiments included in the fit. This method was introduced by Ref. \cite{Hobbs:2019gob}. The study in Ref. \cite{Jing:2023isu} used the methodology in Ref. \cite{Hobbs:2019gob} to calculate the various $L_2$ sensitivities of the experiments in the ATLAS21, CT18, and MSHT global QCD fits. The $L_2$ sensitivities were plotted over $x$ for several $Q$ values and published at \cite{L2website}. Finally, Ref. \cite{Jing:2023isu} created a program to easily calculate and create plots for the $L_2$ sensitivity of any experiment for a given PDF set, called \textbf{L2LHAExplorer}. Similar to the approach in \cite{Jing:2023isu}, I use the L2LHAExplorer program to examine the sensitivities of the ten chosen data sets for CT18, CT18As, and MSHT20. Various contributors provide all of these datasets in xFitter, enabling swift calculations of $L_2$ sensitivities through the extraction of the xFitter output values. It's important to note that CT18/CT18As do not encompass the same sets as MSHT20, and vice versa. Some sets I study are included in both sets, and some are not included in either.

To calculate the $L_2$ sensitivity following the methodology in Ref. \cite{Jing:2023isu}, I presume the total $\chi^2$ to be approximately Gaussian around the total minimum, $\chi^2_0$, within the 68\% confidence level (C.L.). I assume linearity in  $\chi^2$ for each experiment, $\chi^2_E$, within the specified $\chi^2$ range. The tolerance $T^2$, featured in the likelihood probability, can impact the 68\% C.L. range due to the assumed Gaussian nature of $\chi^2$,
\begin{equation}
     P(f)\propto \exp[-\frac{\chi^2(f)-\chi^2_0}{2T^2}].
     \label{eq:LikelihoodProb}
\end{equation}
\noindent A tolerance of $T^2=10$ is applied when constructing the PDF error sets for this study to stay in the parameter region where $\chi_E^2$ are close to being linear.

To generate the $L_2$ sensitivity plots, I tabulated $\chi^2$ contributions and the number of data points from each experiment for each of the PDF error sets using the xFitter framework. I input these data into L2LHAExplorer for plotting the $L_2$ sensitivities, setting the pole charm mass $m_c=1.3$ GeV, bottom mass $m_b=4.75$ GeV, and $\alpha_s(M_Z)=0.118$ for CT18 and CT18As, and $m_c=1.4$ GeV, bottom mass $m_b=4.75$ GeV, and $\alpha_s(M_Z)=0.118$ for MSHT20. These values are taken from the native charm and bottom masses of the respective PDF sets.

Each experiment's $L_2$ sensitivity is calculated by finding the Pearson correlation between $\chi^2_E$ and PDF $f(x, Q)$. In the Hessian representation, the $L_2$ sensitivity of $f$ is given by
\begin{equation}
     S^H_{f,L_2}(E)=\frac{\Vec{\nabla}\chi^2_E\cdot\Vec{\nabla}f}{\delta_Hf}=\delta_H\chi^2_E\times C_H(f,\chi^2_E),
     \label{eq:$L_2$sensitivity}
\end{equation}
\noindent where
\begin{equation}
     C_H(f,\chi^2_E)=\frac{1}{4\delta_Hf\, \delta_H\chi^2_E}\sum^D_{i=1}(f_{+i}-f_{-i})(\chi^2_{E,+i}-\chi^2_{E,-i})
     \label{eq:PearsonCorrelation}
\end{equation}
\noindent is the cosine of correlation between $f$ and $\chi^2_E$ with $D$ as the number of eigenvector pairs. $\delta_Hf$ and $\delta_H\chi^2_E$ are the $1\sigma$ uncertainties for $f$ and the $\chi^2_E$ \cite{Hobbs:2019gob}.

Due to the direct proportionality between $S^H_{f,L_2}$ and $C_H$, the correlation serves as a reliable indicator of whether experiments prefer a smaller or larger PDF at given $x$ and $Q$ values. A positive correlation, and therefore a positive $S^H_{f,L_2}$, indicates that the $\chi^2_E$ will increase when $f$ increases by 1$\sigma$, and will decrease otherwise, resulting in the experiment favoring a smaller PDF to minimize the $\chi^2_E$. However, anticorrelated $f$ and $\chi^2_E$ indicate a negative $L_2$ sensitivity, favoring a positive pull on the PDF. The larger the magnitude of $S^H_{f,L_2}$ is, the stronger the experiment pulls on the PDF to reduce its $\chi^2_E$. 

\section{Results}
\label{Results}

\begin{figure}
    \centering
    \includegraphics[width=\columnwidth]{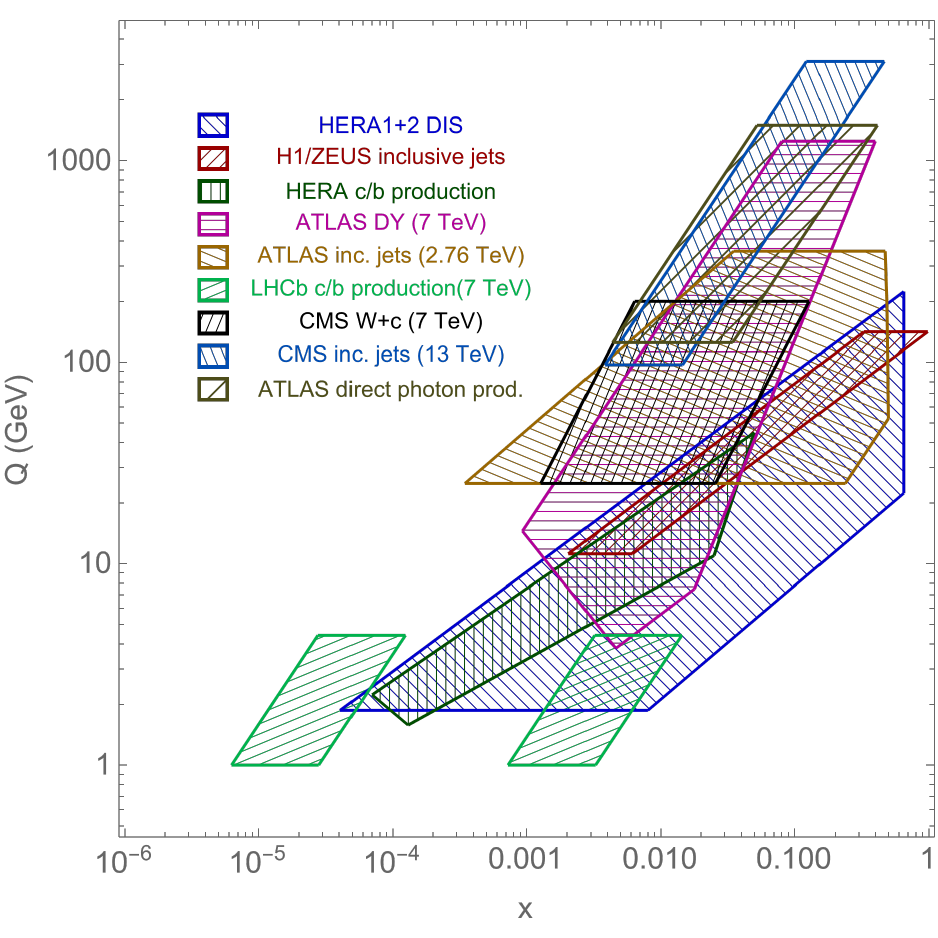}
    \caption{Approximate $x$ and $Q$ (GeV) distribution of the data points from various experiments included in this study.}
    \label{fig:L2_xQ2}
\end{figure}

%make table full width of paper
\begin{table*}
    \centering
    \begin{tabular}{c|c|c|c}
          & CT18 & CT18As & MSHT20 \\
         \hline
         ATLAS direct $\gamma$ production 8 and 13 TeV \cite{ATLAS:2019drj} & $\times$ & $\times$ & $\times$ \\
         ATLAS DY 7 TeV \cite{ATLAS:2013xny,ATLAS:2014ape} & $\times$ & $\times$ & \checkmark \\
         ATLAS inc. jet production 2.76 TeV \cite{ATLAS:2013pbc} & $\times$ &  $\times$ & \checkmark \\
         CMS inc. jet production 13 TeV \cite{CMS:2021yzl} & $\times$ & $\times$ & $\times$ \\
         CMS $W$+$c$ production 7 TeV \cite{CMS:2013wql} & $\times$ & $\times$ & $\times$  \\
         H1+ZEUS $c$ and $b$ production \cite{H1:2018flt} & $\times$ & $\times$ &\checkmark \\
         H1 jet production \cite{H1:2007xjj,H1:2010mgp} & & & \\
         HERA I+II combined DIS \cite{H1:2015ubc} & \checkmark & \checkmark & \checkmark \\
         LHCb $c$ and $b$ production 7 TeV \cite{LHCb:2013xam,LHCb:2013vjr} & $\times$ & $\times$ & $\times$ \\
         ZEUS jet production \cite{ZEUS:2002nms,ZEUS:2006xvn,ZEUS:2010vyw} & & & \\
    \end{tabular}
    \caption{A table indicating which datasets examined in this study are included in the CT18, CT18As, and MSHT20 PDF global analyses.}
    \label{tab:datasets-in-PDFs}
\end{table*}

I focused on 10 experiments in this study, all of which are implemented in the xFitter package [cf. Tab.~\ref{tab:datasets-in-PDFs}]. Table~\ref{tab:datasets-in-PDFs} indicates which experiments were included in the CT18, CT18As, and MSHT20 PDF sets during their respective global analyses. Fig.~\ref{fig:L2_xQ2} shows their approximate kinematic coverages in the $x$-$Q$ plane. Concise overviews of each experiment are provided in the following sections. For detailed information on the experiments, please refer to the relevant papers in the bibliography.

\begin{figure*}
    \centering
    \includegraphics[width=\columnwidth]{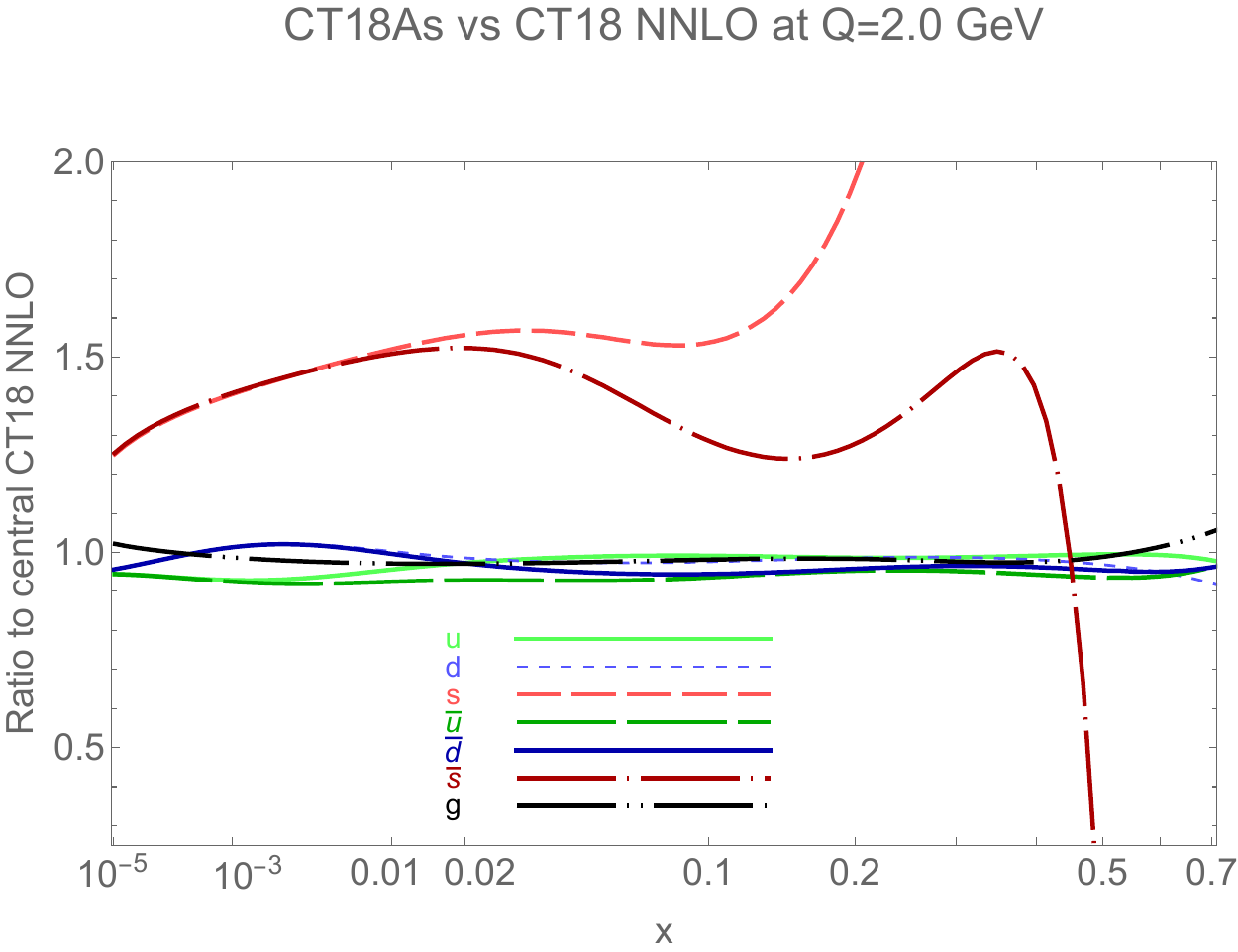}
    \includegraphics[width=\columnwidth]{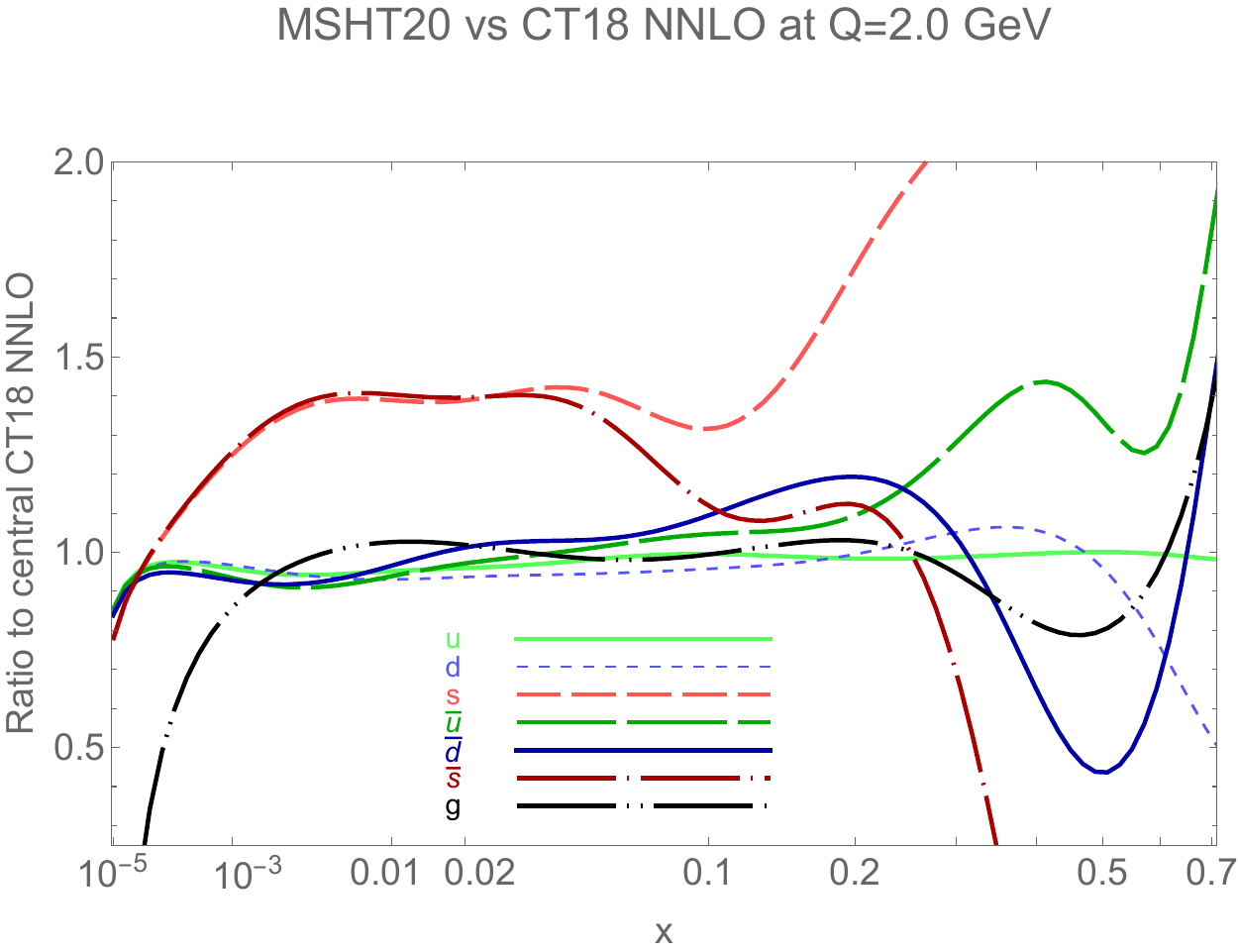}
    \includegraphics[width=\columnwidth]{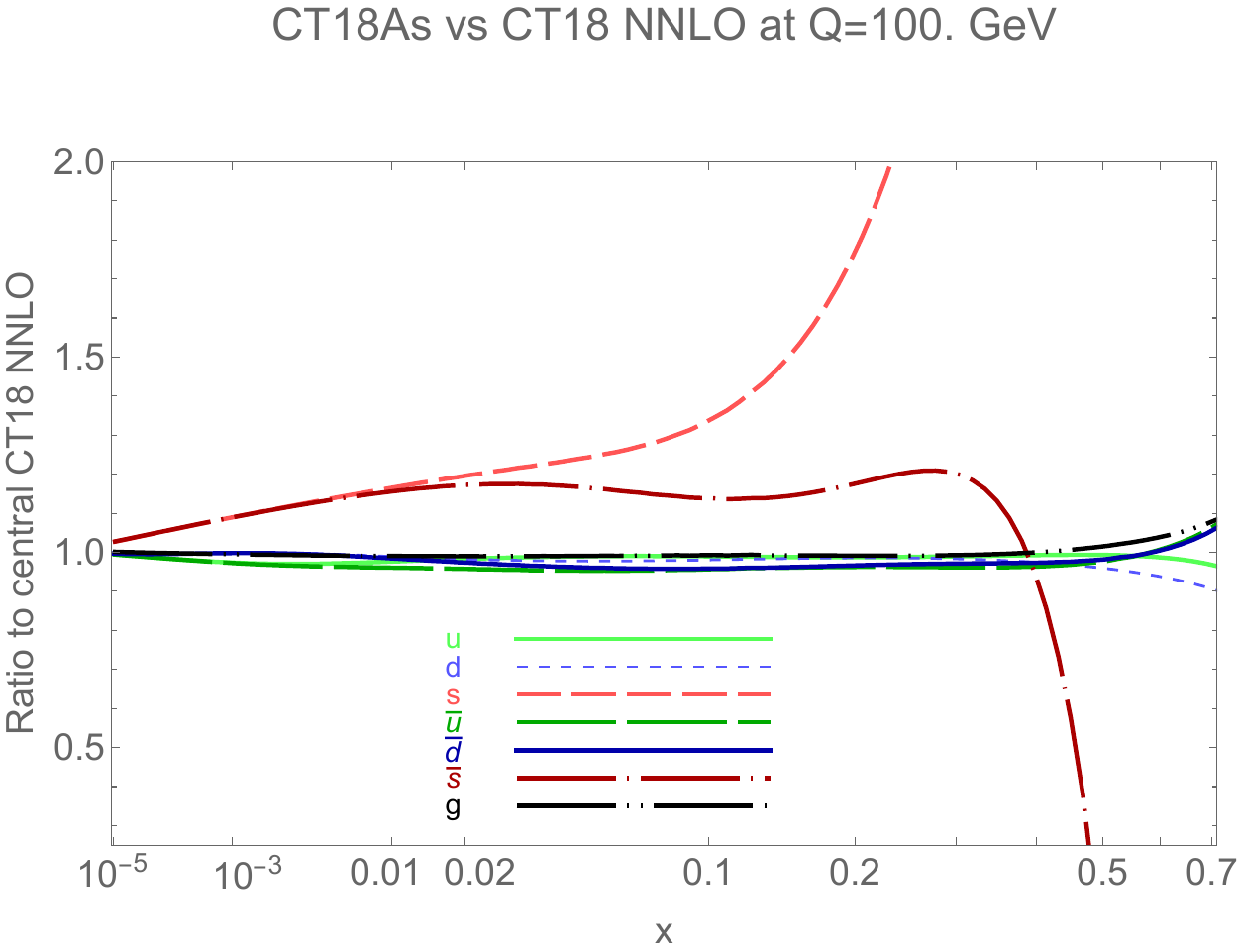}
    \includegraphics[width=\columnwidth]{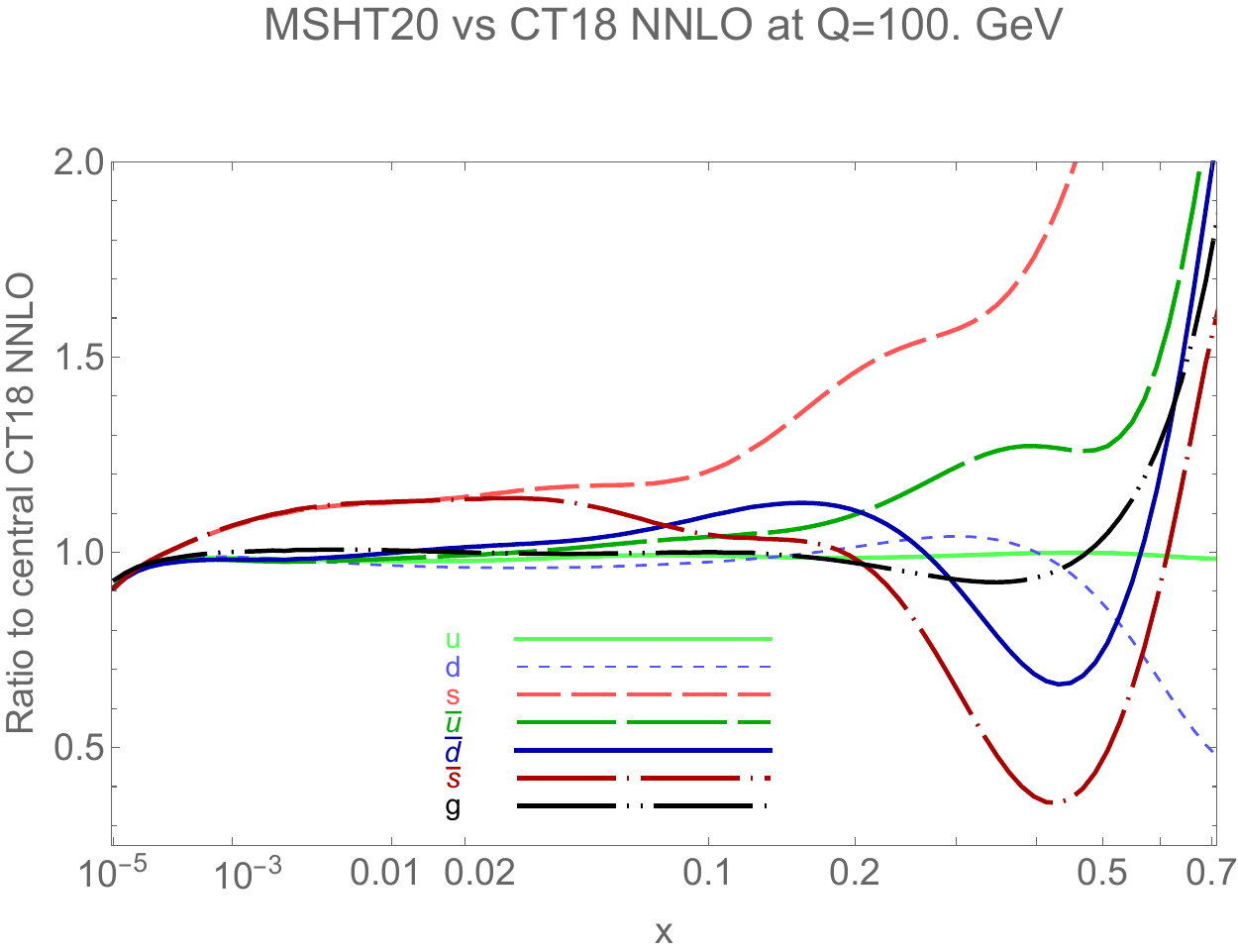}
    \caption{CT18As (left) and MSHT20 (right) vs CT18 NNLO central PDFs at $Q=2$ and 100 GeV.}
    \label{fig:PDFRatios}
\end{figure*}

As mentioned in Sec.~\ref{Introduction}, I analyze sensitivities to CT18, CT18As, and MSHT20 PDFs at NNLO with a tolerance of $T^2=10$. I obtained these PDF sets from the authors of Ref.~\cite{Jing:2023isu}. Before examining the sensitivities, I examine the differences between these PDF sets, which have been explored in much greater detail in \cite{Amoroso:2022eow,PDF4LHCWorkingGroup:2022cjn}. Figure~\ref{fig:PDFRatios} plots the ratios of CT18As and MSHT20 to CT18. In the left plot, $s$ and $\bar{s}$ PDFs are larger in CT18As than in CT18, and the asymmetry of the strange PDFs becomes apparent at $x\sim0.01$. The other flavors in the left plot (Fig.~\ref{fig:PDFRatios}) are approximately equal for both CT18 and CT18As. Experiments that are less sensitive to $s$ and $\bar{s}$ yield comparable $L_2$ sensitivities as a result of the very similar PDFs. Consequently, plots of CT18As sensitivities are omitted for the experiments insensitive to $s/\bar{s}$. Examination of the right plot in Fig. 2 shows a similar trend in the MSHT20 vs CT18 ratios for the $s$ and $\bar{s}$ PDFs. The MSHT20 PDFs for $u, d$, and their antiparticles are comparable with CT18 in the low-$x$ region ($x\lesssim0.1$). The MSHT20 gluon PDF trends downwards from CT18 in the region $x\lesssim10^{-3}$, where the MSHT20 gluon eventually becomes negative around $x\lesssim10^{-5}$. I will discuss the consequence of the negative gluon PDF in this region at $Q\sim 1$ GeV in Sec. 3.3.

\subsection{HERA I+II inclusive deep-inelastic scattering and charm and beauty production}
\label{sec:HERA}

\subsubsection{Inclusive DIS}
\label{sec:HERADIS}

\begin{figure*}
    \centering
    \includegraphics[width=1.\columnwidth]{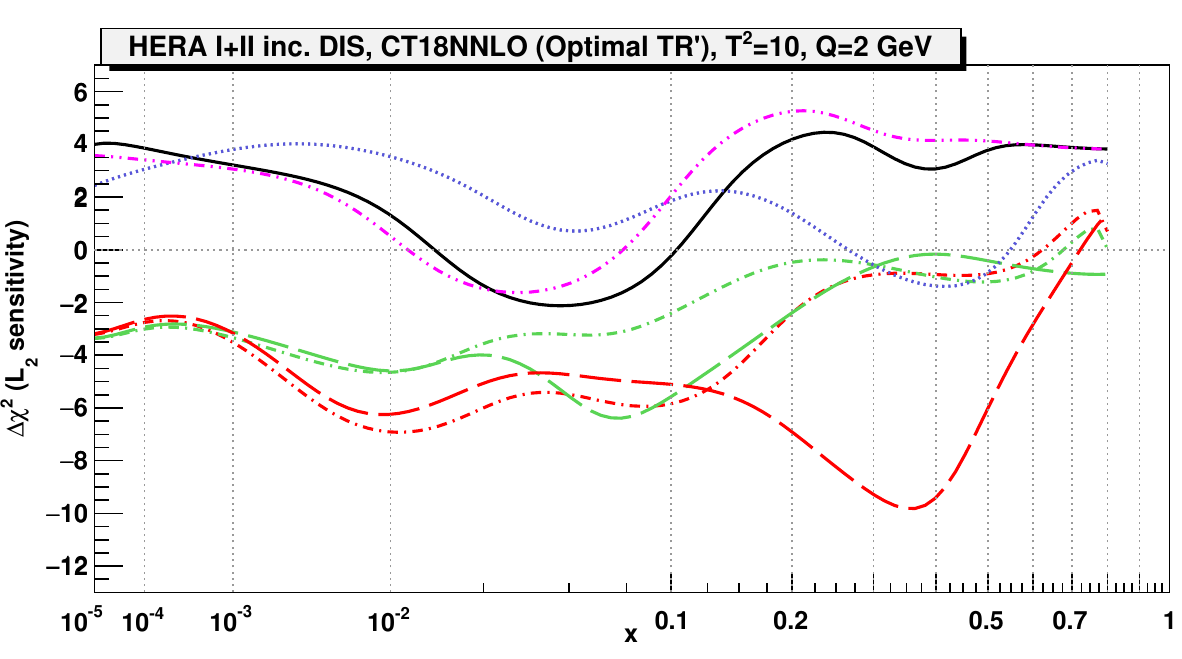}
    \includegraphics[width=1.\columnwidth]{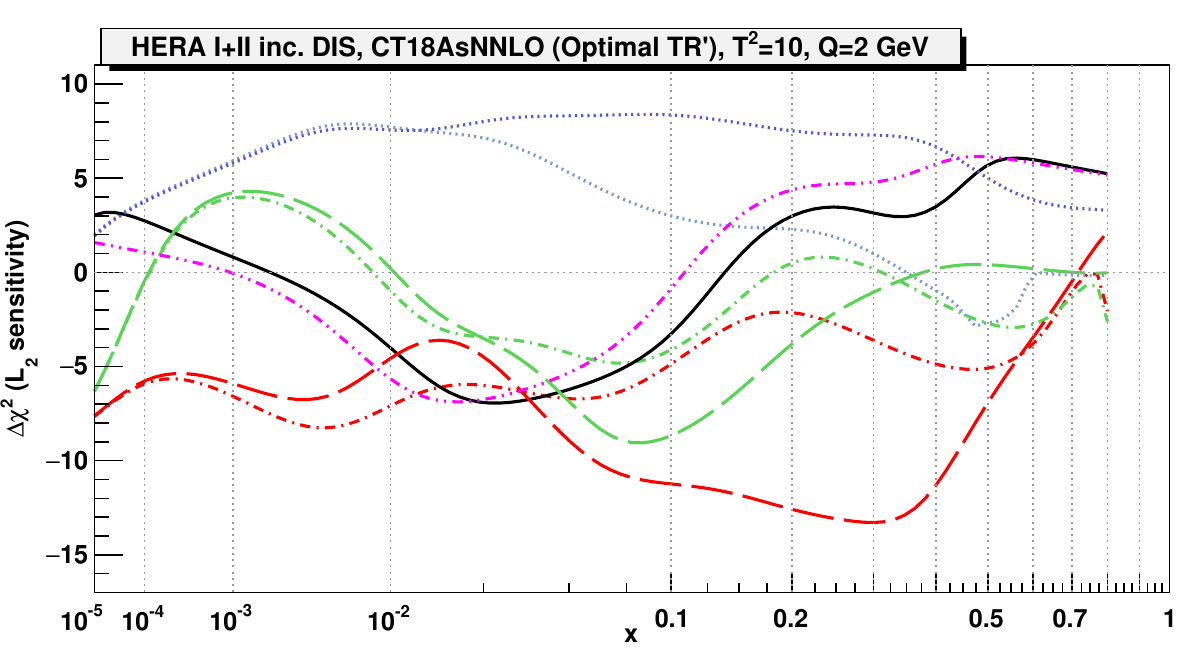}
    \includegraphics[width=1.\columnwidth]{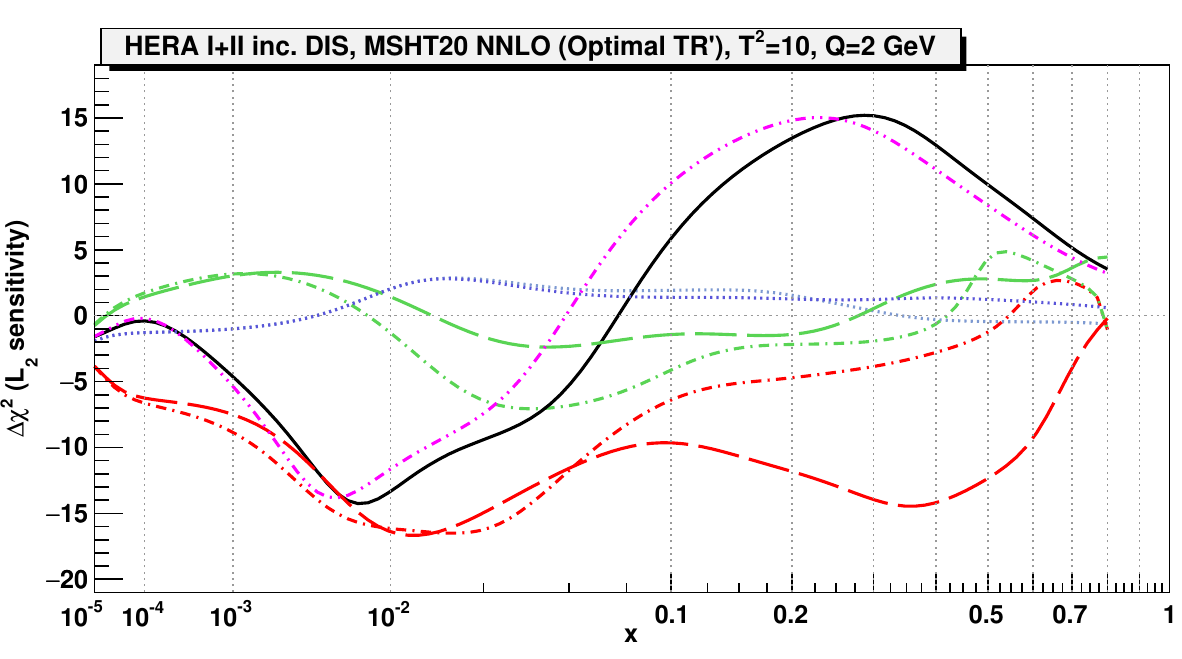}
    \includegraphics[width=0.35\columnwidth]{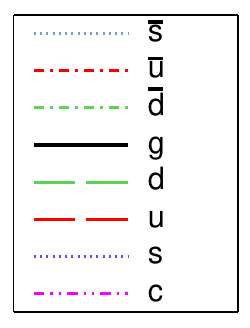}
    \caption{$L_2$ sensitivity for HERA I+II combined inclusive DIS \cite{H1:2015ubc} using the CT18, CT18As, and MSHT20 NNLO PDF sets with $T^2=10$. For this process, xFitter uses the optimal TR' scheme to evaluate contributions from heavy quarks.}
    
    \label{fig:HERA I+II}
\end{figure*}

I validated my procedure using the HERA combined DIS experiment as a reference \cite{H1:2015ubc}. This experiment was studied with the same settings in Ref. \cite{Jing:2023isu}, with the sensitivities computed using the native CTEQ-TEA and MSHT codes and plotted on \cite{L2website}. I utilized the inclusive DIS cross sections from H1 and ZEUS collaborations for $e^\pm p$ scattering at zero beam polarization in both neutral current (NC) and charged current (CC) cases. Fig.~\ref{fig:L2_xQ2} shows the kinematic coverage which spans approximately $5\times 10^{-5}\lesssim x \lesssim 0.7$ and $2\lesssim Q\lesssim 200$ GeV. The measured observable - the reduced DIS cross section - is most sensitive to the quark PDFs, as reflected by the large sensitivities to the up, down, and their respective antiquark PDFs, $u$, $d$, $\bar{u}$, and $\bar{d}$.

For all three PDF sets, the HERA I+II combined inclusive DIS experiment (Fig.~\ref{fig:HERA I+II}) shows the highest sensitivity among all examined experiments, as expected due to its large number of data points (1145 points). The sensitivity to the strangeness PDF, $s(x,Q)$, has the largest magnitude in CT18As out of all three PDF sets and is positive for all $x$ values. CT18 and MSHT sensitivities to $s(x,Q)$ both are slightly negative in specific regions. Another contrast between CT18 and CT18As was observed in the sensitivities to $d$ and $\bar{d}$. The CT18 case favored smaller $d$ and $\bar{d}$ sensitivities, while the CT18As case preferred larger $d$ and $\bar{d}$ sensitivities at small $x$.

\begin{figure*}
    \centering    \includegraphics[width=1.0\columnwidth]{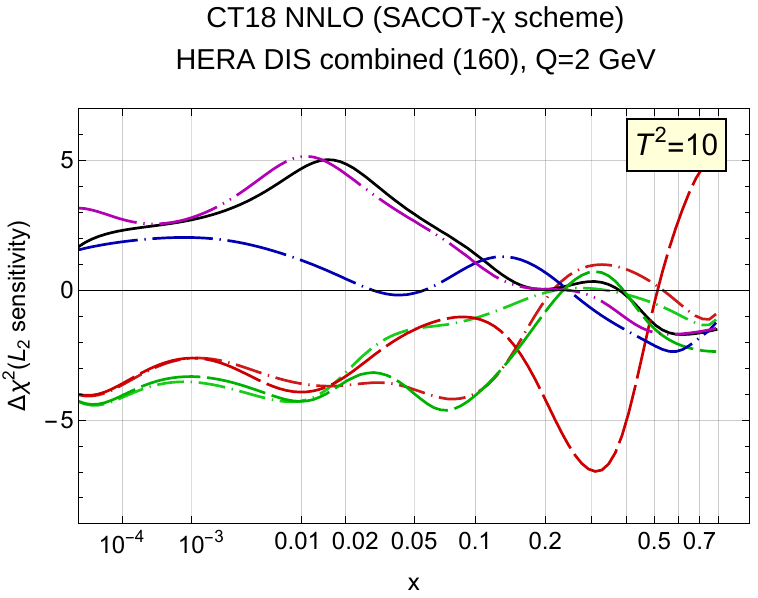}    \includegraphics[width=1.0\columnwidth]{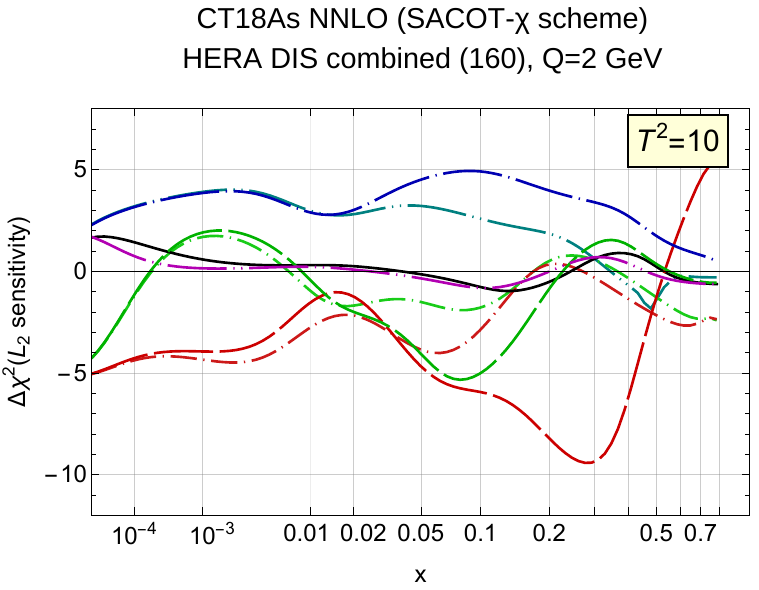}    \includegraphics[width=1.15\columnwidth]{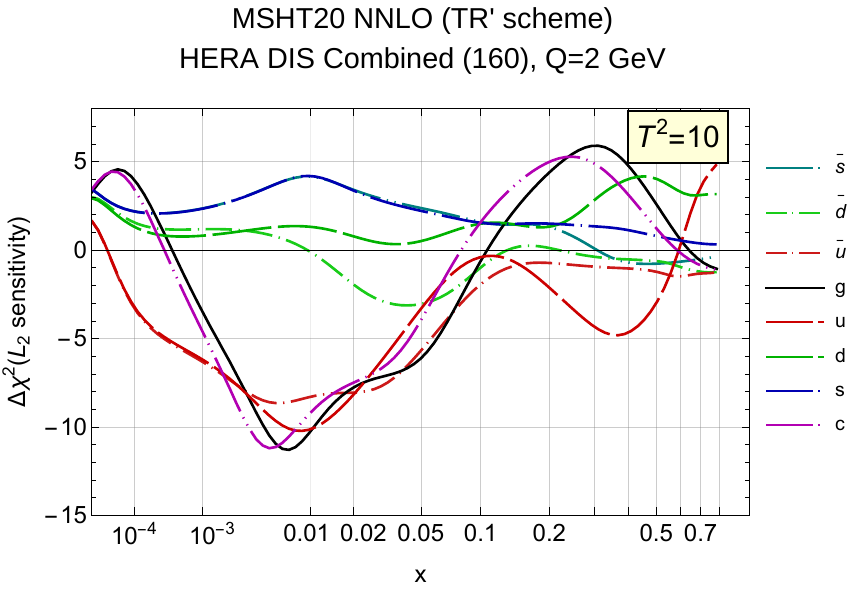}
    \caption{$L_2$ sensitivity from \cite{L2website} for HERA I+II combined inclusive DIS \cite{H1:2015ubc} using the CT18, CT18As, and MSHT20 NNLO PDF sets with $T^2=10$. The heavy-quark scheme used by each group is listed for each of the three PDF sets.}
    \label{fig:HERADISmetapdfL2}
\end{figure*}

The $L_2$ sensitivities for the combined HERA DIS data set obtained with the native CT18 and MSHT20 codes have been already examined in \cite{Jing:2023isu} and plotted in \cite{L2website}. I reproduce these sensitivities from \cite{L2website} in Fig.~\ref{fig:HERADISmetapdfL2}. By comparing the sensitivities in Figs.~\ref{fig:HERA I+II} and \ref{fig:HERADISmetapdfL2}, we learn about the differences in the calculations for the inclusive DIS data set in the CT18, MSHT20, and xFitter codes. The CTEQ-TEA group uses the SACOT-$\chi$ scheme \cite{Hou:2019efy} for the CT18 and CT18As PDF sets, which differs from the MSHT and xFitter groups' utilization of the Thorne-Roberts GM-VFNS modified heavy quark scheme (TR') \cite{Bailey:2020ooq}, potentially causing the slight distinction between my HERA I+II DIS $L_2$ sensitivities for CT18 and those in Ref. \cite{Jing:2023isu}. However, analyzing the MSHT20 plots depicted in Fig.~\ref{fig:HERA I+II} and~\ref{fig:HERADISmetapdfL2} indicates notable disparities in the $L_2$ sensitivities, despite the utilization of the same heavy-quark schemes in both scenarios. For MSHT20 in Fig.~\ref{fig:HERA I+II}, the $L_2$ sensitivities for range approximately from $-15$ to $15$, whereas in the lower subplot of Fig.~\ref{fig:HERADISmetapdfL2}, this range varies between -10 and 5. I, therefore, identify non-negligible differences (compared to $\Delta T^2=10$) in the sensitivities for HERA I+II production for MSHT20 PDFs in the xFitter and MSHT codes using the same TR' scheme. The source of the differences can be investigated in a future study. It can be caused by additional factors, such as the different definitions of $\chi^2$ adopted in the xFitter and MSHT frameworks.
%The difference in the MSHT20 plots strongly supports that the variations between Fig.~\ref{fig:HERA I+II} and \ref{fig:HERADISmetapdfL2} are not solely due to the difference of heavy quark schemes. The reason for these differences must be looked into further studies outside of this one. There are a few aspects to consider when determining the cause of such differences. For example, the definition of $chi^2$ may differ in the global analysis codes and xFitter, or the treatment of certain physical constants in both cases may lead to such differences.}

In Fig.~\ref{fig:HERA I+II} we see another distinctive feature that persists across subsequent plots shown in this paper. The charm sensitivity closely mimics the gluon behavior, showing only a slight magnitude difference across all $x$ and $Q$ values. This pattern is explained by $g\rightarrow{q\bar{q}}$ production. A gluon inside of a hadron can split into a $c\bar{c}$ pair during the collision process. However, the timing of this process determines which parton originating from the hadron enters the hard interaction region. When the splitting occurs before the interaction region, either the $c$ or $\bar{c}$ particle enters the hard interaction region just like any other initial-state parton and is associated with its PDF, which is called flavor excitation. As a result, the sensitivities to the charm and gluon are very similar across all figures.

\subsubsection{Charm and beauty production at HERA}
\label{sec:HERAcb}

\begin{figure*}
    \centering
    \includegraphics[width=1.0\columnwidth]{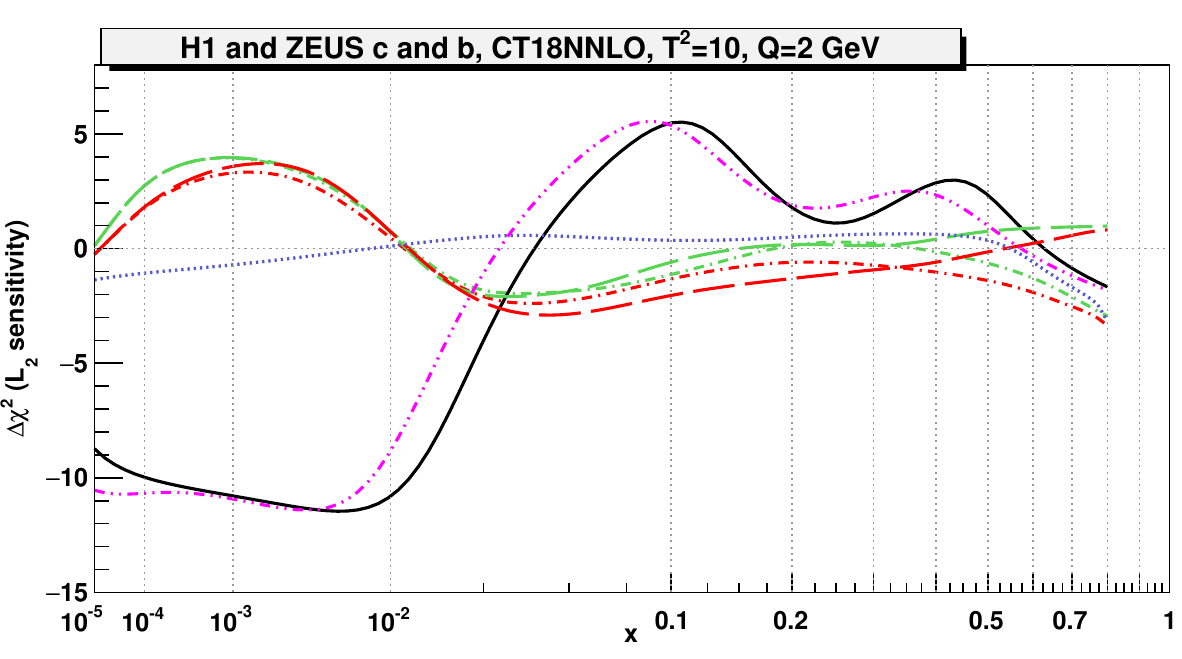}
    \def\big{\includegraphics[width=1.0\columnwidth]{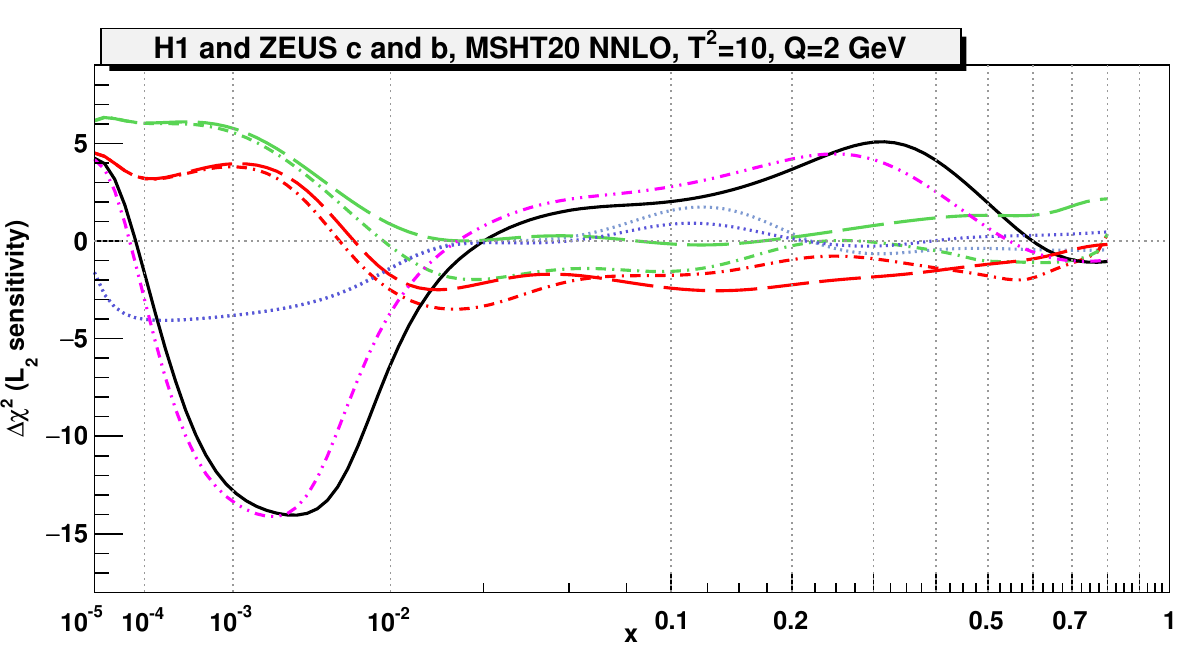}}
    \def\little{\includegraphics[width=0.2\columnwidth]{figs/L2_study/ssbarlegend.pdf}}
    \stackinset{r}{8pt}{b}{15pt}{\little}{\big}
    \caption{$L_2$ sensitivity for H1+ZEUS combined charm and beauty production \cite{H1:2018flt} using the CT18, and MSHT20 NNLO PDF sets with $T^2=10$.}
    \label{fig:H1+ZEUScb}
\end{figure*}

The combined charm and bottom production data set from HERA \cite{H1:2018flt} covers the kinematic range $2\lesssim Q \lesssim 40$ GeV and $7 \times 10^{-5} \lesssim x \lesssim 2 \times 10^{-2}$ [cf. Fig.~\ref{fig:L2_xQ2}]. These data sets allow for the measurement of the gluon PDF due to the leading hard-scattering reaction studied here being $\gamma^* g\rightarrow q\bar{q}$, where $q$ is the heavy quark.

Despite the data set comprising 74 data points — contrasting with HERA I+II inclusive DIS data set with $\sim$1000 points — H1+ZEUS combined charm and beauty experiment exhibits substantial gluon/charm sensitivities compared to the other experiments (Fig.~\ref{fig:H1+ZEUScb}). For all three PDF sets, there is notable negative sensitivity to the gluon in the low-$x$ region, followed by positive sensitivity in the large-$x$ region. The key distinction between CT18 and MSHT sets is that CT18 sensitivities to HERA $c$ and $b$ production is largely negative for small $x$, while the MSHT one is negative for $10^{-4}\lesssim x \lesssim 2\times 10^{-2}$, but becomes positive for $x\lesssim10^{-4}$. This could be a possible result of the difference in the MSHT gluon PDF from CT18 in that region [cf. Fig.~\ref{fig:PDFRatios}]. For $x\lesssim10^{-4}$ and $x\gtrsim0.1$, the MSHT20 light quarks deviate from the CT18 central fit as well, contributing to differences in the sensitivities in those regions.

\subsection{CMS W+c production (\(\sqrt{s}=7\) TeV)}
\label{sec:CMSW+c}

\begin{figure*}
    \centering
    \includegraphics[width=1.0\columnwidth]{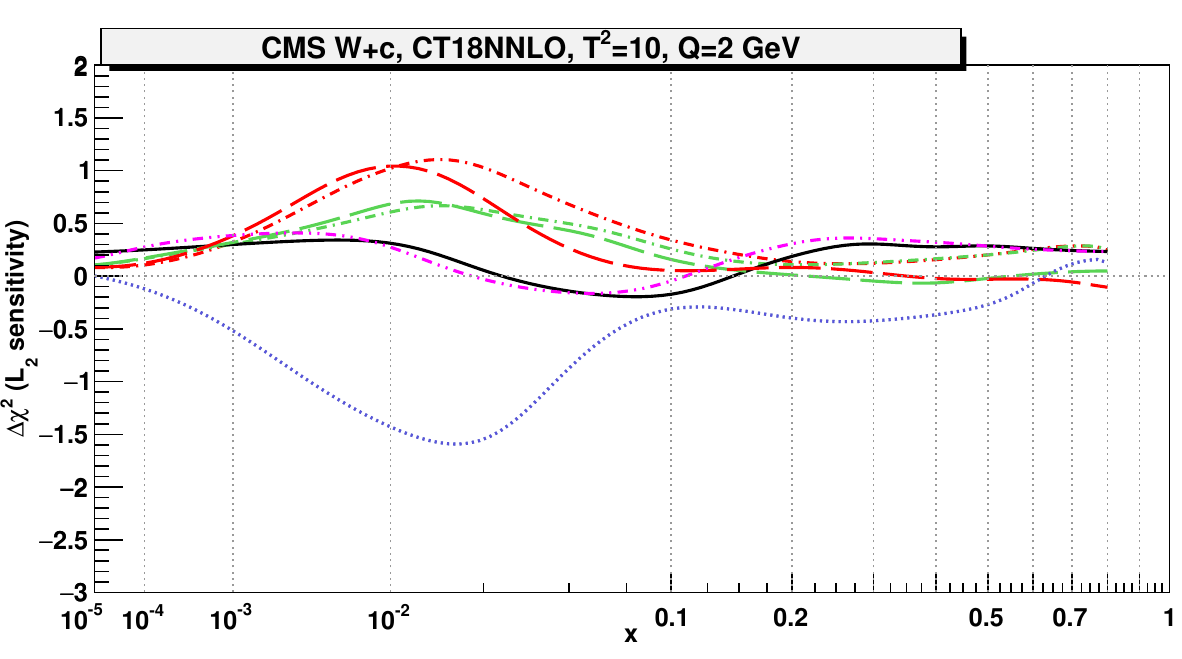}
    \includegraphics[width=1.0\columnwidth]{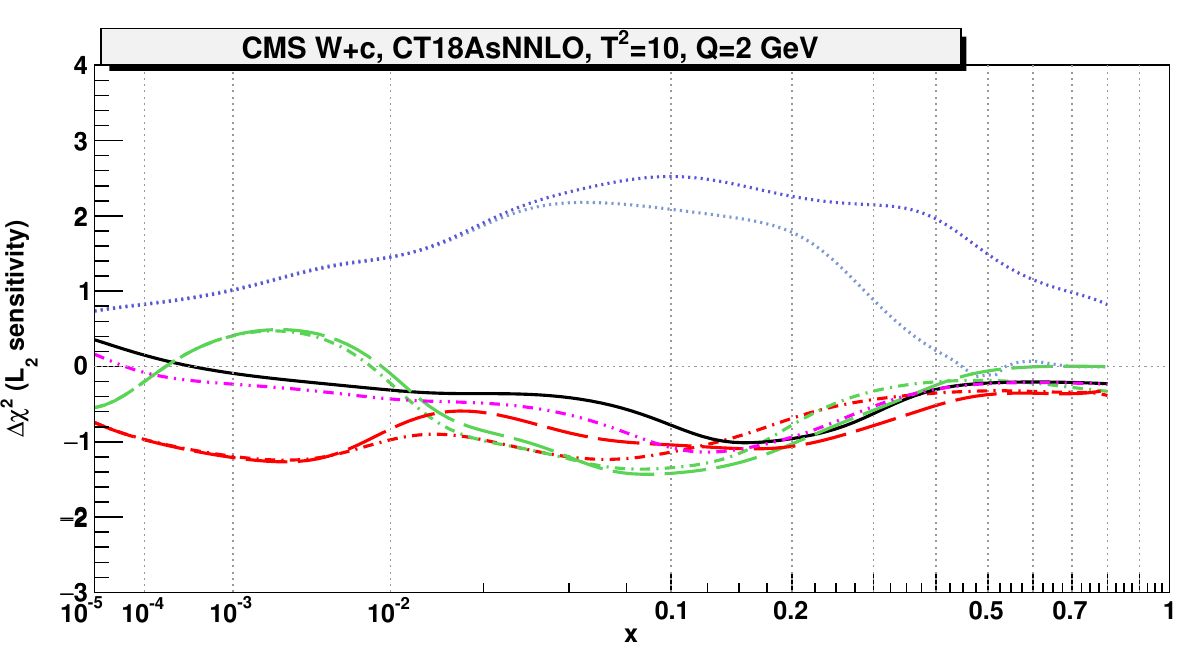}
    \def\big{\includegraphics[width=1.0\columnwidth]{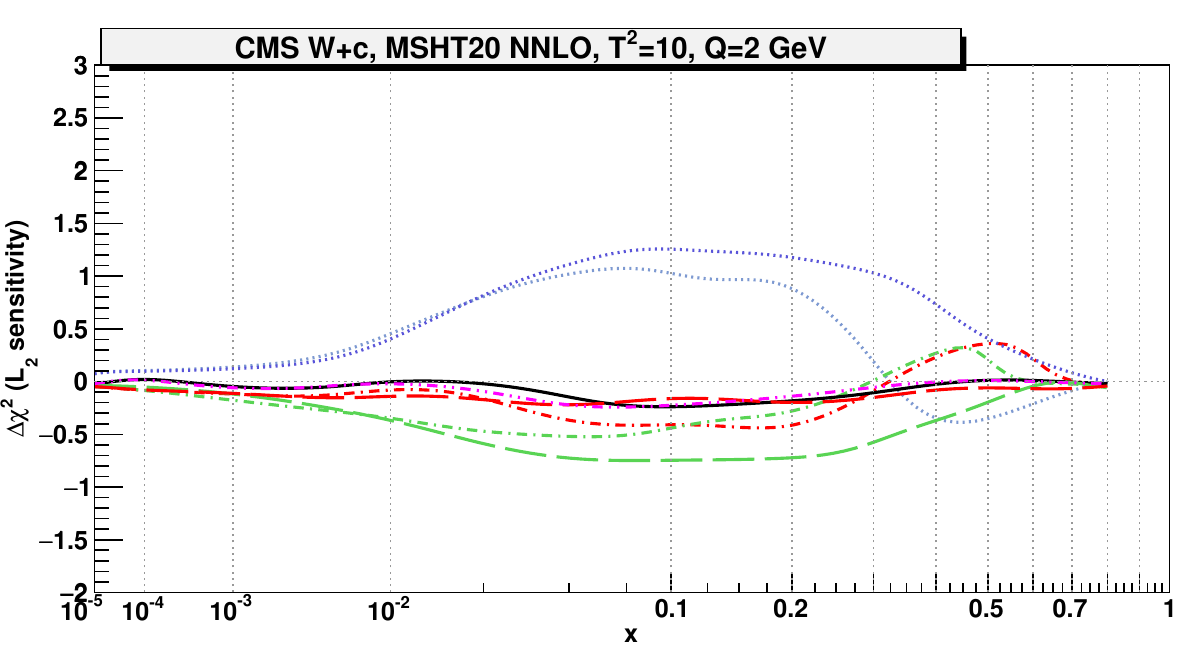}}
    \def\little{\includegraphics[width=0.2\columnwidth]{figs/L2_study/ssbarlegend.pdf}}
    \stackinset{l}{22pt}{t}{15pt}{\little}{\big}
    \caption{$L_2$ sensitivity for CMS 7 TeV W+c production \cite{CMS:2013wql} using the CT18, CT18As, and MSHT20 NNLO PDF sets with $T^2=10$.}
    \label{fig:CMSWplusc}
\end{figure*}

The specific reactions studied in this experiment \cite{CMS:2013wql} are $pp \rightarrow W^++\bar{c}+X$ and $pp \rightarrow W^-+c+X$ at $\sqrt{s}=7$ TeV. The primary contribution stems from $s/\bar{s}$, with a slightly lesser contribution from $d/\bar{d}$ due to the $W$ boson favoring interactions with same-generation particles \cite{Czakon:2021jvx}. This dominance arises from the most probable $Wc$ production channel: $s+g\rightarrow W^-+c$ and $\bar{s}+g \rightarrow W^++\bar{c}$. For the DIS experiments, I evaluated hard cross sections, $\hat{\sigma}$, at NNLO in $\alpha_s$. Meanwhile, for the CMS experiment, I calculated $\hat{\sigma}$ at NLO in $\alpha_s$ given the absence of an NNLO calculation. This is also true for the following experiments in this study, where calculations were performed at NLO.

Figure~\ref{fig:CMSWplusc} shows the following trends among all three PDF sets. The first one is that the $s$ sensitivity ranges between -1.5 and 2.5 across all PDF sets. This is a relatively small sensitivity compared to the ones of the dominant experiments. Consequently, CMS W+c production would impose a weaker pull on the strangeness within a global fit. However, the $s$ PDF varies among the three PDF sets, resulting in either mostly negative or positive sensitivities for all $x$. The second one is that the sensitivities to all other flavors are considerably lower than to $s$: this experiment is most sensitive to the $s$ PDF, as mentioned above. CT18As, which is the asymmetric strangeness version of CT18A, includes the ATLAS W/Z experiments, unlike the CT18 set. The ATLAS W/Z experiment is found to prefer a larger $s$ than W+c production when included in the global analysis. Given the lower CT18 $s$ compared to CT18As [cf. Fig~\ref{fig:PDFRatios}], CMS W+c production is inclined to increase the CT18 strangeness PDF and reduce the CT18As counterpart. Thus, W+c production at CMS favors an enhancement in strangeness compared to CT18, in concordance with other DY experiments at the LHC.

\subsection{LHCb \(c\) and \(b\) production ($\sqrt{s}=7$ TeV)}
\label{sec:LHCBresults}

Numerous measurements were performed at LHCb for pro-
mpt charm and $b$ quark production \cite{LHCb:2013xam,LHCb:2013vjr} at $\sqrt{s}=7$ TeV. The experiment measured $D/B$ mesons within $c/b$ tagged jets. In LHCb $c$ and $b$ production, the lowest-order process is $gg\rightarrow(c\rightarrow D)X$ or $gg\rightarrow(b\rightarrow B)X$, with radiative contributions to the hard cross section $\hat \sigma$ in $\alpha_s$ included up to NLO in the xFitter calculation. The approximate kinematic ranges measured are $6\times10^{-6}\lesssim x \lesssim 10^{-4}$ for charm and $7\times10^{-4}\lesssim x\lesssim 0.015$ for beauty, at $1\lesssim Q \lesssim 5$ GeV [cf. Fig.~\ref{fig:L2_xQ2}].

\begin{figure*}
    \centering
    \includegraphics[width=1.0\columnwidth]{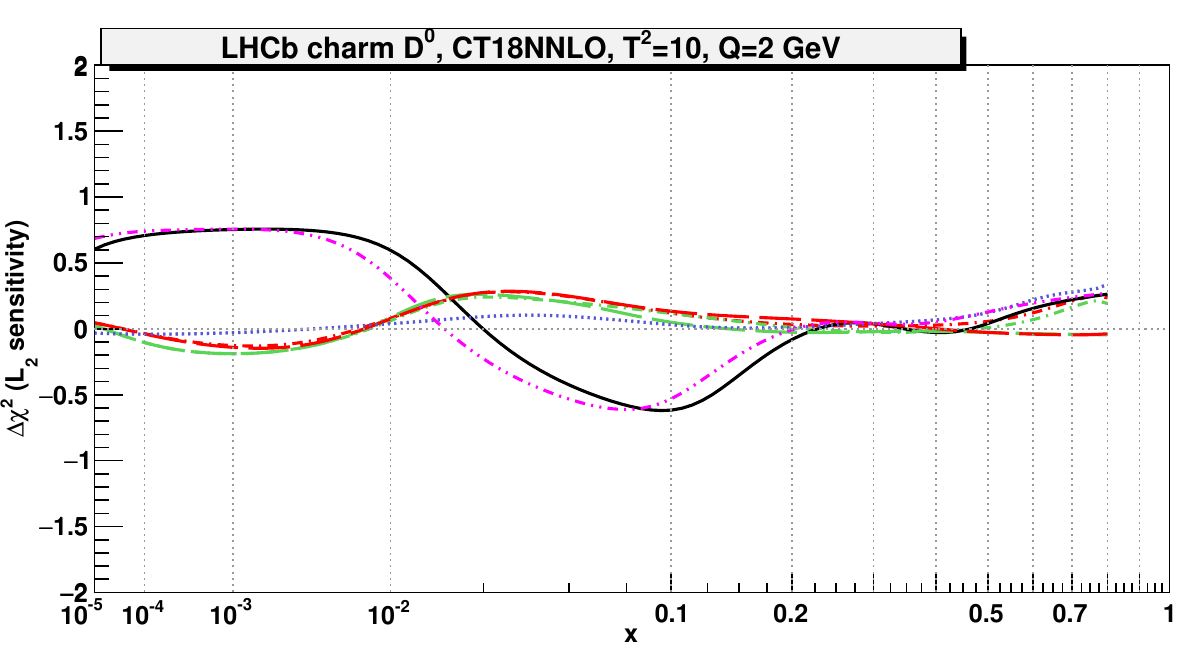}
    \def\big{\includegraphics[width=1.0\columnwidth]{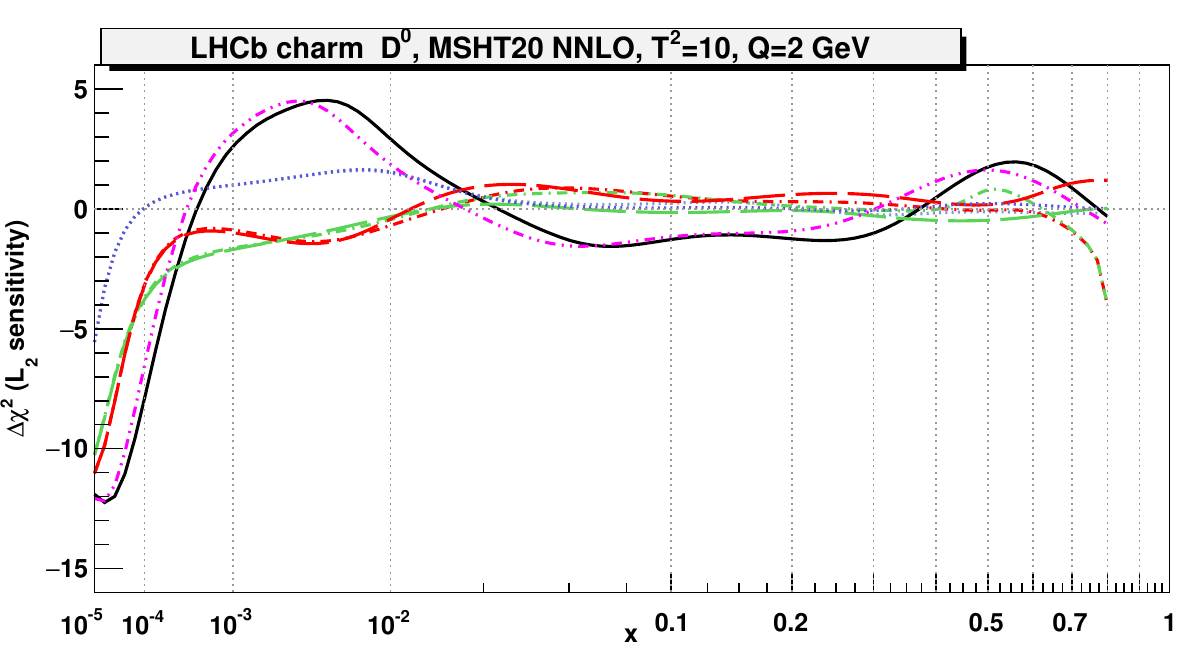}}
    \def\little{\includegraphics[width=0.2\columnwidth]{figs/L2_study/ssbarlegend.pdf}}
    \stackinset{r}{10pt}{b}{15pt}{\little}{\big}
    \caption{$L_2$ sensitivity for LHCb $D^0$ production \cite{LHCb:2013xam} using the CT18, and MSHT20 NNLO PDF sets with $T^2=10$, with the cut $p_T^{meson}\geq2$ GeV.}
    \label{fig:LHCbD0pTcut}
\end{figure*}

For LHCb, in Fig.~\ref{fig:L2_xQ2} I estimate the approximate range of $x$ using the rapidity $y$ of the reconstructed meson, center-of-mass energy $\sqrt{s}$, and transverse momentum of the meson jet $p_T^{meson}$,
\begin{equation}
    x_{1,2}\approx\frac{p_T^{meson}}{s}e^{\pm y_{meson}}.
    \label{eq:DYbjroken}
\end{equation}
\noindent $x_1$ and $x_2$ are the order-of-magnitude estimates of dominant momentum fraction for proton beams 1 and 2. The default QCD scales are set to $p_T^{meson}$ in my analysis. According to Eq. (\ref{eq:DYbjroken}), $p_T^{meson}\leq1$ GeV results in data points at $x\lesssim 10^{-5}$ and QCD scales no more than 1 GeV, i.e. in the kinematic region where perturbative QCD becomes unstable, as seen in Fig.~\ref{fig:L2_xQ2}. Therefore, I explore sensitivities for $p_T^{meson} \geq 2$ GeV to reduce the dependence of $S^H_{f,L_2}$ on $x<10^{-4}$ and scales of 1 GeV or less. \(p_{T, \text{min}}^{meson}\) cuts of 1 and 3 GeV were also explored. However, a 1 GeV cut failed to sufficiently remove \(x \sim 10^{-5}\) data points, while a 3 GeV cut overly eliminated points without significant sensitivity improvement. A \(p_T^{meson} \geq 2\) GeV cut effectively eliminated minimum data points, while also eliminating the problematic small-$x$ influence. I show the respective sensitivities in Fig.~\ref{fig:LHCbD0pTcut}. By default, xFitter does not impose any such cuts and includes all data points from the experimental data. 

These cuts were necessary to avoid nonperturbative regions and also thereforestabilized the behavior of the MSHT20 PDF set at such low scales. In the MSHT20 set, the gluon PDF becomes negative at $x<10^{-4}$ and $Q<2$ GeV, as implied in Fig.~\ref{fig:PDFRatios}. Without \(p_T^{meson}\) cuts, erratic \(L_2\) sensitivities for MSHT PDFs occurred at small \(x\) possibly due to the negative gluon PDF. At larger $Q$, the MSHT20 gluon becomes positive at small $x$, and as a result the respective $L_2$ sensitivities behave reasonably with the cut $p_T^{meson}\geq 2$ GeV in the right Fig.~\ref{fig:LHCbD0pTcut}. Nevertheless, the corresponding $L_2$ sensitivity indicates the preference for a larger gluon at $x>10^{-3}$ than in the MSHT20 PDF set, see the right Fig.~\ref{fig:LHCbD0pTcut}. Notably, CT18 and CT18As sensitivities did not exhibit this issue, showing consistent sensitivities except for a subtle magnitude increase.

In addition to exploring various \(p_T^{meson}\) cuts, the study assessed the impact of adjusting the factorization scale (\(\mu_F\)) and the renormalization scale (\(\mu_R\)). xFitter allows users to modify these scales by a user-defined factor in the input files. I investigated the consequences of independently setting \(\mu_{R/F}\) to 0.5\(\mu_{R/F}^{(0)}\) and 2\(\mu_{R/F}^{(0)}\) for the LHCb experiment, where $\mu_{R/F}^{(0)}$ is the initial scale used.

Setting \(\mu_{R/F}\) to 2\(\mu_{R/F}^{(0)}\) slightly decreased the sensitivities of LHCb measurements, while at 0.5\(\mu_{R/F}^{(0)}\), sensitivities marginally increased. These outcomes, not presented in this paper, do not provide additional insights beyond the subtle changes in magnitude. With the cut, the data prefer about the same gluon for CT18 and a larger gluon for MSHT at $x<10^{-4}$.

% \begin{figure*}
%     \centering
%     \includegraphics[width=1.0\columnwidth]{figs/L2_study/CT18NNT210-LHCb-cb-D0_Q2_0.pdf}
%     \includegraphics[width=1.0\columnwidth]{figs/L2_study/MSHT20nnlo_as118_set_new__July21_T210_pdf-LHCb-cb-D0_Q2_0.pdf}
%     \caption{$L_2$ sensitivity for LHCb $D^0$ production experiment \cite{LHCb:2013xam} using the CT18, and MSHT20 NNLO PDF sets with $T^2=10$, with no $p_T^{meson}$ cut.}
%     \label{fig:LHCbD0}
% \end{figure*}

\subsection{CMS inclusive jet production 13 TeV}

%\NOTE{Include 7 and 8 TeV CMS L2 sensitivities and make a comment comparing 13 TeV to the two. Explore the compatibility of 13 TeV with 7 or 8.}

The CMS group measured the double-differential cross sections for inclusive jets at $\sqrt{s}=13$ TeV \cite{CMS:2021yzl}. Unlike the previous plots examined, the 13 TeV CMS jet production dataset sensitivities were examined for PDFs evaluated at scale $Q=100$ GeV. With a much higher center-of-mass energy, the lowest $Q$ measured is 97 GeV, and the highest is $\sim$1200 GeV.

\begin{figure*}
    \centering
    \includegraphics[width=1.0\columnwidth]{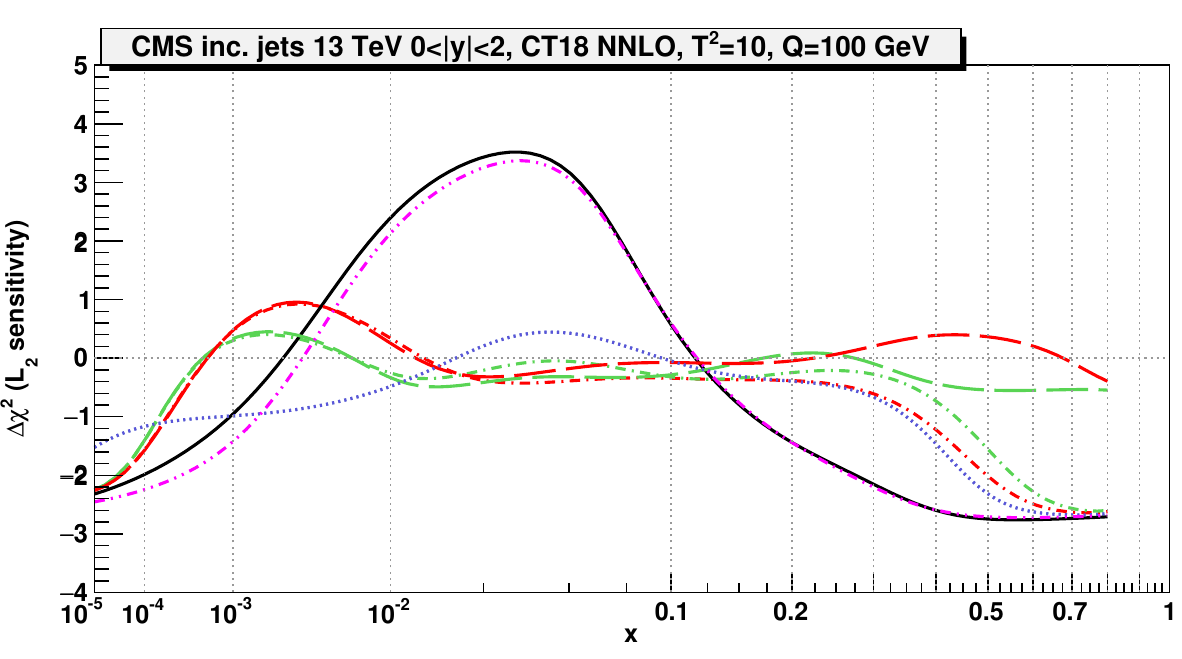}
    \def\big{\includegraphics[width=1.0\columnwidth]{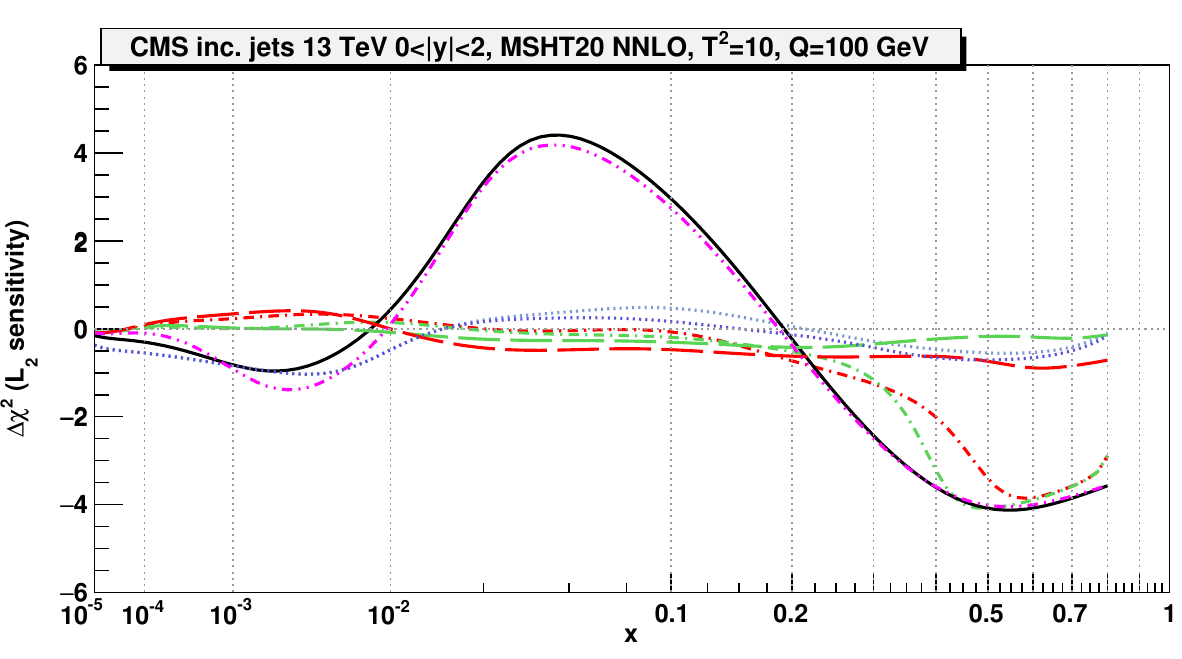}}
    \def\little{\includegraphics[width=0.15\columnwidth]{figs/L2_study/ssbarlegend.pdf}}
    \stackinset{r}{5pt}{t}{15pt}{\little}{\big}
    \caption{$L_2$ sensitivity for CMS inclusive jet production 13 TeV with $0<|y|<2$ \cite{CMS:2021yzl} using the CT18, and MSHT20 NNLO PDF sets with $T^2=10$.}
    \label{fig:cms13tev}
\end{figure*}

Figure~\ref{fig:cms13tev} shows that for $10^{-2}<x<0.1$, the CMS 13 TeV jet experiment favors a smaller gluon PDF for all three PDF sets, while preferring a larger gluon for $x>0.3$. A notable distinction between CT18/CT18As and MSHT20 is that CT18/CT18As favor a larger gluon distribution for $x<10^{-4}$, whereas MSHT20 does not. In the small $x$ range, it's also seen that the experiment prefers larger quark PDFs for CT18/CT18As, while it is not particularly sensitive in the MSHT20 set. A smaller $\bar{u}$ and $\bar{d}$ are both preferred for CT18/CT18As and MSHT20 at large $x$, but only a smaller $s$ is preferred in CT18/CT18As at large $x$.

\begin{figure*}
    \centering
    \includegraphics[width=1.0\columnwidth]{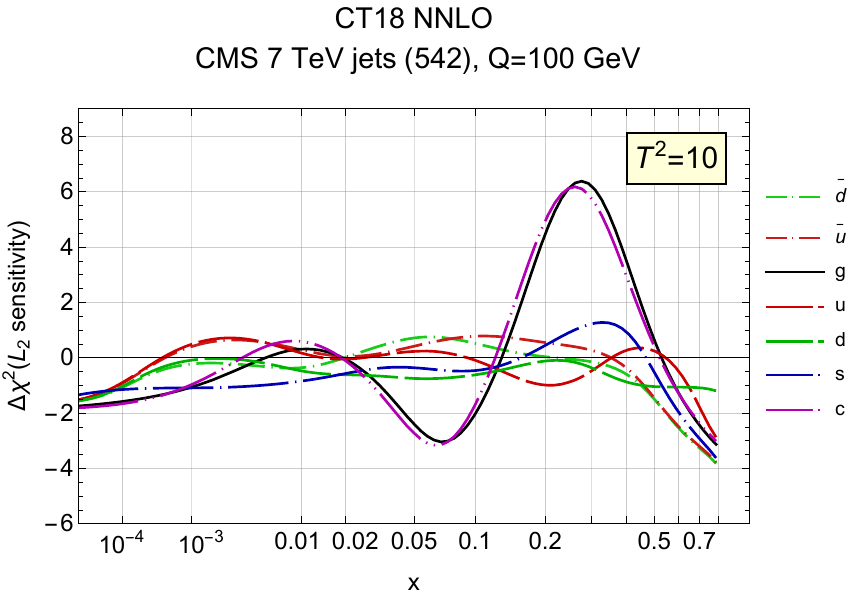}
    \includegraphics[width=1.0\columnwidth]{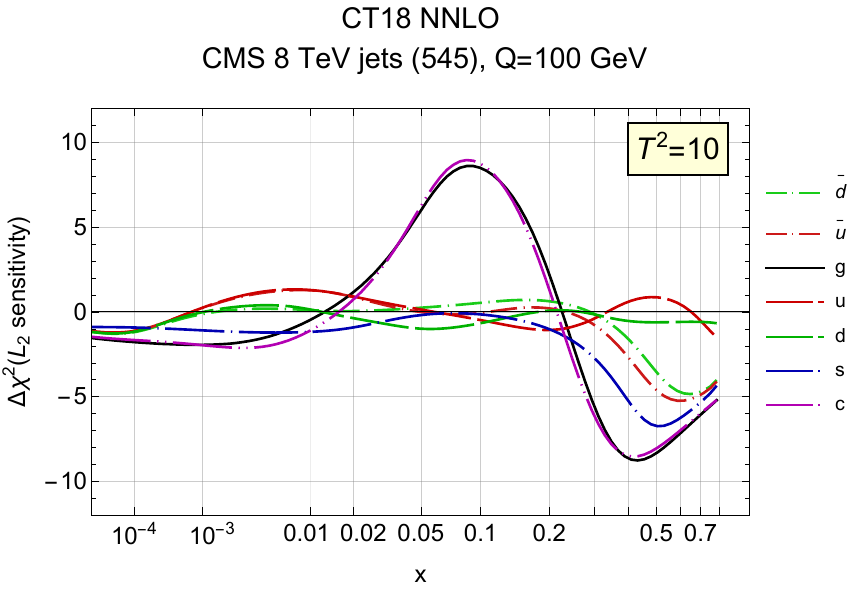}
    \includegraphics[width=1.0\columnwidth]{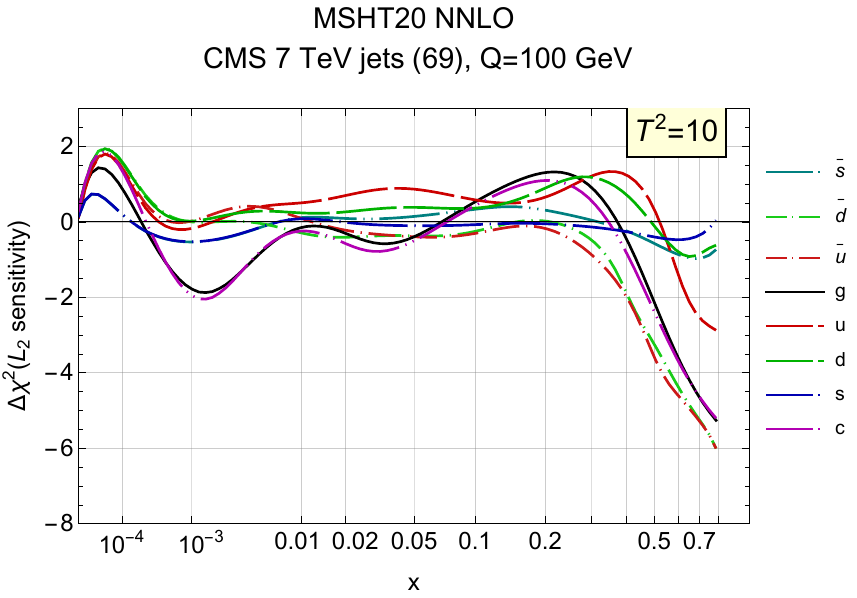}
    \includegraphics[width=1.0\columnwidth]{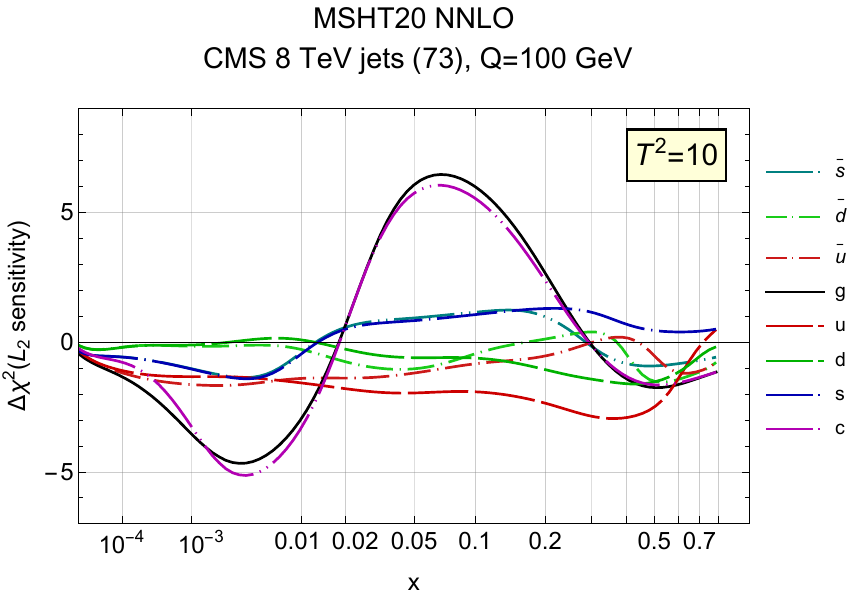}
    \caption{$L_2$ sensitivity for CMS inclusive jet production 7 and 8 TeV using CT18 and MSHT20 calculated by \cite{L2website}.}
    \label{fig:jingcms13tev}
\end{figure*}

When examining the potential impact of the CMS 13 TeV inclusive jet dataset, compatibility with the 7 and 8 TeV datasets must also be considered. The peak of the sensitivity plots in Fig.~\ref{fig:cms13tev} is expected to roughly shift by a factor of $\frac{\sqrt{s}}{13 TeV}$, where $\sqrt{s}$ is either 7 or 8 TeV. Figure~\ref{fig:jingcms13tev} reveals the shift between 8 TeV and 13 TeV is as expected, but not for 7 and 13 TeV. This may indicate some underlying factors that are not apparent in this study, prompting further potential investigation into the compatibility of the CMS jets data at 7 TeV with the 13 TeV dataset as the 7 TeV sensitivities do not reflect similar behaviors to those exhibited in the 13 TeV data.

\subsection{Less sensitive experiments}
\label{sec:lesssensitive}

The following experiments have considerably lower sensitivities compared to the previous experiments examined above. I document their sensitivities for possible future comparisons.

\subsubsection{ZEUS and H1 inclusive jet production}
\begin{figure*}
    \centering
    \includegraphics[width=1.0\columnwidth]{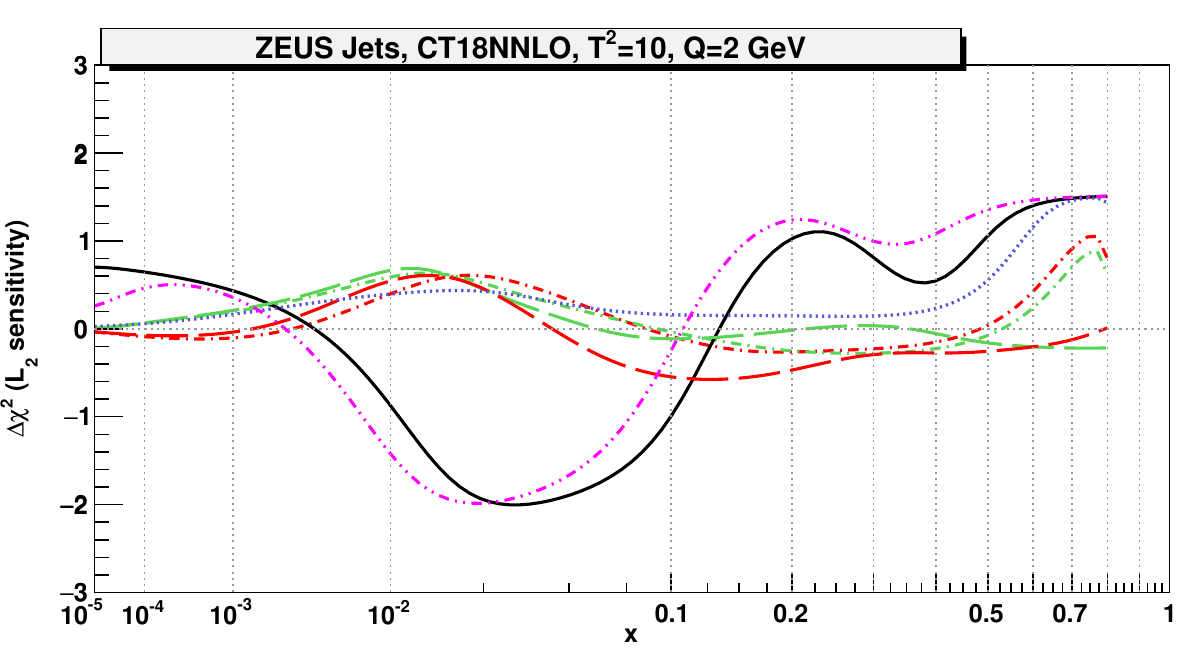}
    \def\big{\includegraphics[width=1.0\columnwidth]{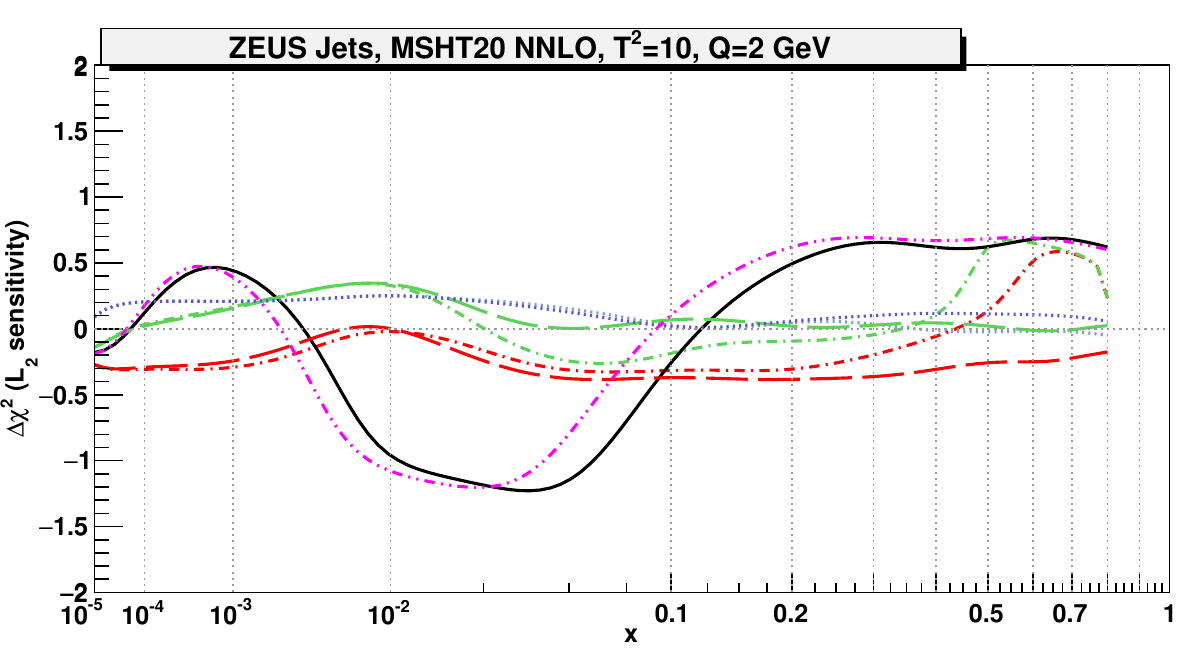}}
    \def\little{\includegraphics[width=0.15\columnwidth]{figs/L2_study/ssbarlegend.pdf}}
    \stackinset{l}{110pt}{t}{18pt}{\little}{\big}
    \caption{$L_2$ sensitivity for ZEUS jet production \cite{ZEUS:2002nms,ZEUS:2006xvn,ZEUS:2010vyw} using the CT18, and MSHT20 NNLO PDF sets with $T^2=10$.}
    \label{fig:ZEUSJets}
\end{figure*}

The dominant reaction for the NC DIS ZEUS jet production \cite{ZEUS:2002nms,ZEUS:2006xvn,ZEUS:2010vyw} is QCD Compton scattering ($\gamma^* q\rightarrow q g$) or boson-gluon fusion ($\gamma^* g\rightarrow q\bar{q}$) at LO. Figure~\ref{fig:ZEUSJets} shows a larger gluon sensitivity compared to the light quarks. The measured kinematic range is $11\lesssim Q \lesssim 100$ GeV. and $2\times 10^{-3} \lesssim x \lesssim 0.6$ [cf. Fig.~\ref{fig:L2_xQ2}].

\begin{figure*}
    \centering
    \includegraphics[width=1.0\columnwidth]{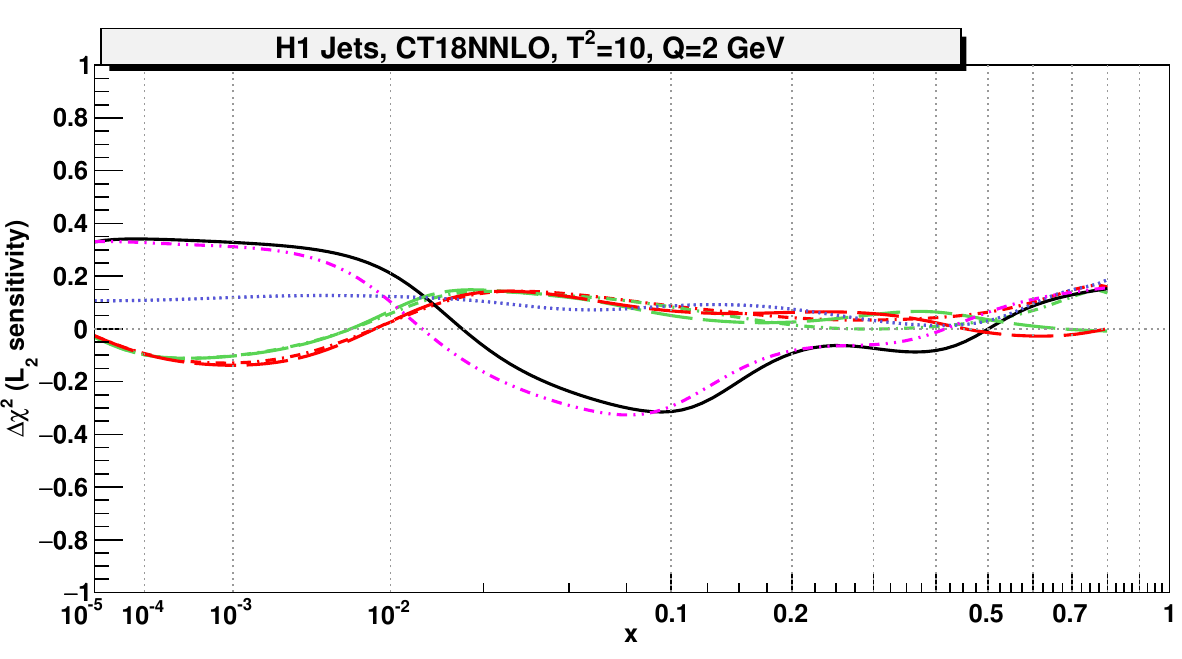}
    \def\big{\includegraphics[width=1.0\columnwidth]{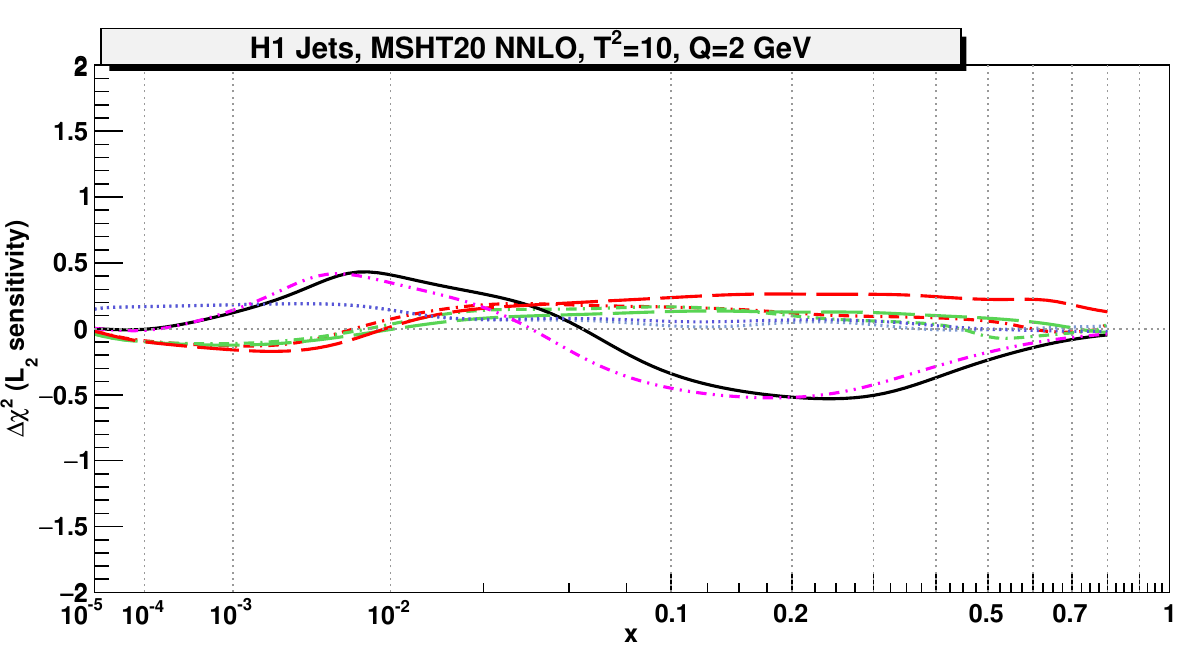}}
    \def\little{\includegraphics[width=0.15\columnwidth]{figs/L2_study/ssbarlegend.pdf}}
    \stackinset{r}{2pt}{t}{10pt}{\little}{\big}
    \caption{$L_2$ sensitivity for H1 jet production \cite{H1:2007xjj,H1:2010mgp} using the CT18, and MSHT20 NNLO PDF sets with $T^2=10$.}
    \label{fig:H1Jets}
\end{figure*}

H1 jet production experiment \cite{H1:2007xjj,H1:2010mgp} is similarly most sensitive to the gluon PDF inside the proton in two approximate $Q$ ranges: high momentum transfer ($Q\gtrsim12$ GeV) and low $Q$ ($2\lesssim Q \lesssim 10$ GeV). Figure~\ref{fig:H1Jets} demonstrates comparable sensitivities for all quarks in MSHT20 and CT18, all of marginal amplitude. The H1 sensitivity to the gluon is more subdued than for ZEUS: despite the H1 and ZEUS jet production experiments being very similar they produce dissimilar results, for the gluon sensitivity. This may be possibly attributed to the difference in $E_T^{beam}$ cuts chosen in each experiment.

\subsubsection{ATLAS DY 7 TeV}
%\NOTE{Refer to two papers mentioned in referee comments. Emphasize that the plots include low and high mass invariant masses. Specify mass ranges. High mass region includes NLO EW correction, and the N3LO corrections can be either positive or negative depending on the kinematic ranges and the PDF behavior, which needs to be verified by future calculation.}

\begin{figure*}
    \centering
    \includegraphics[width=1.0\columnwidth]{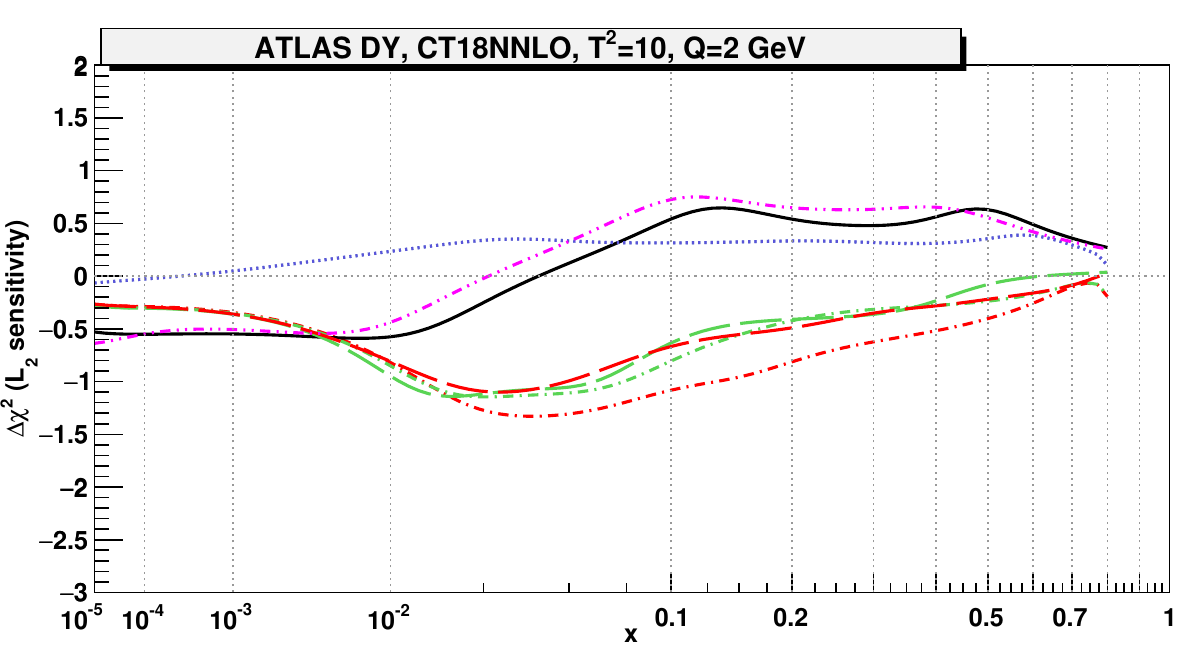}
    \includegraphics[width=1.0\columnwidth]{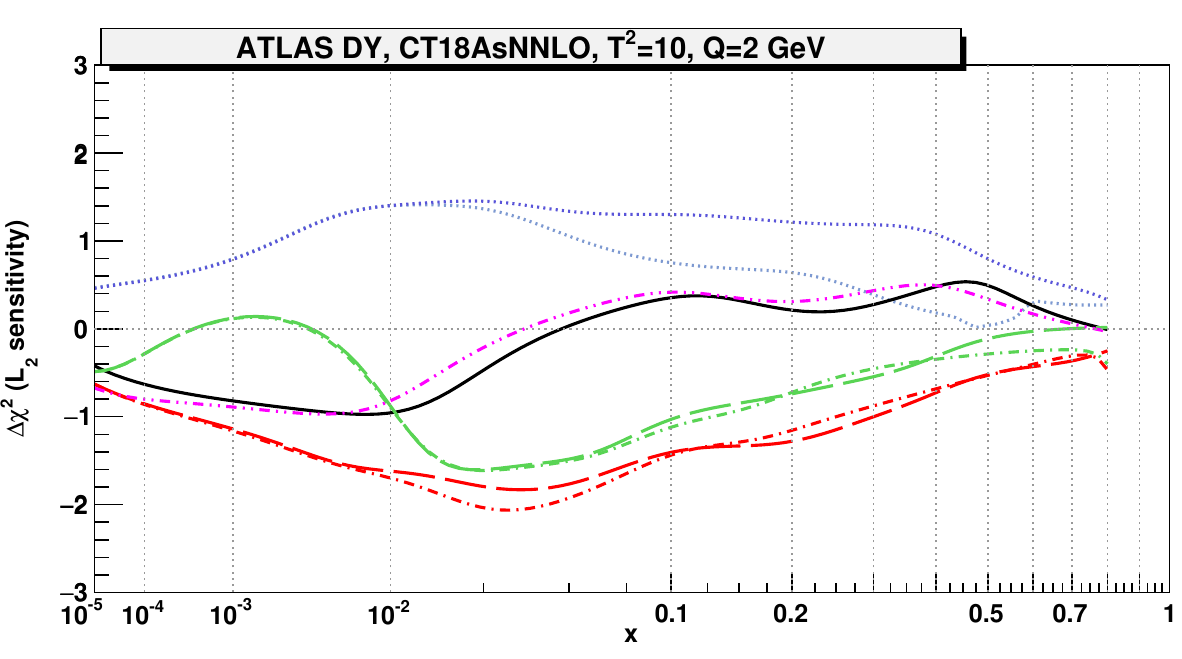}
    \def\big{\includegraphics[width=1.0\columnwidth]{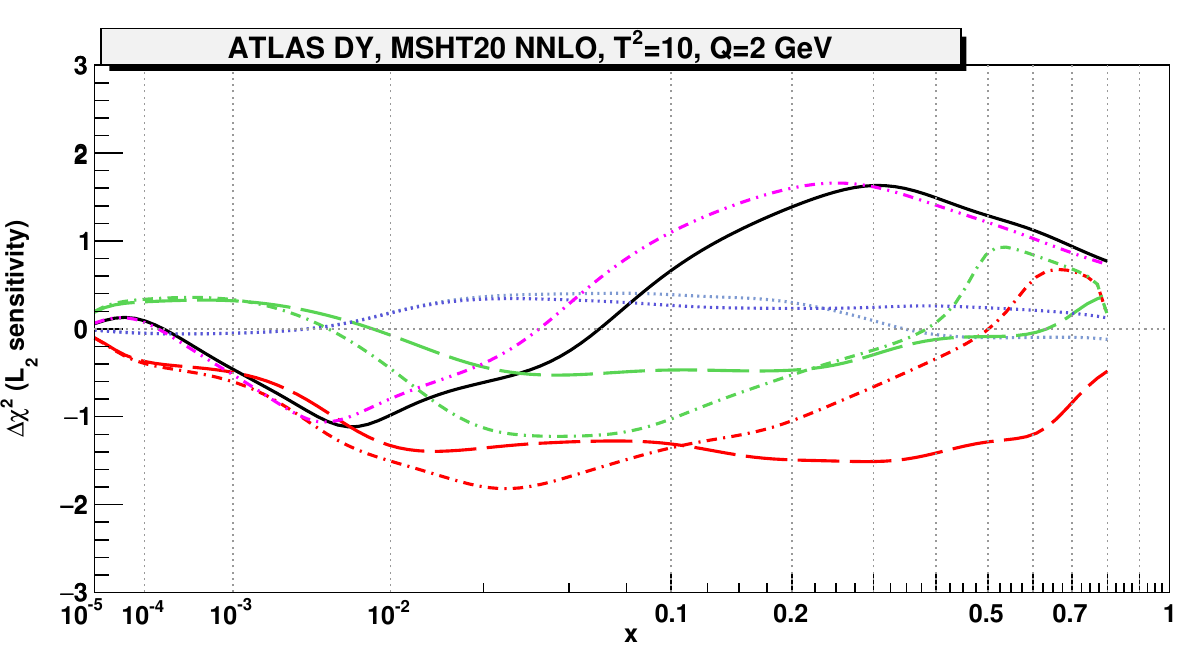}}
    \def\little{\includegraphics[width=0.15\columnwidth]{figs/L2_study/ssbarlegend.pdf}}
    \stackinset{l}{70pt}{t}{20pt}{\little}{\big}
    \caption{$L_2$ sensitivity for ATLAS DY pair production \cite{ATLAS:2013xny,ATLAS:2014ape} using the CT18, and MSHT20 NNLO PDF sets with $T^2=10$.}
    \label{fig:ATLASDY}
\end{figure*}

The ATLAS group measured the Drell-Yan differential cross section of $Z/\gamma^*\rightarrow \ell^+\ell^-$ at $\sqrt{s}=7$ TeV \cite{ATLAS:2013xny,ATLAS:2014ape}. Figure~\ref{fig:ATLASDY} shows the respective sensitivities. $\ell=e$ for high invariant lepton mass. The final results are a linear combination of the $\ell=e$ and $\ell=\mu$ channels for low invariant lepton mass. The dominant reaction contributing to the measurements is $q\bar{q}\rightarrow \ell^+\ell^-$. However, when $Q$ is low, the reactions including gluons in the initial state ($gg\rightarrow\ell^+\ell^-$ and $gq\rightarrow\ell^+\ell^-$) are enhanced due to the large enhancement of the gluon PDF at low $Q$ and $x$. The data provided by the experiment is separated into two ranges of the invariant lepton mass: for the low mass range, $12$ GeV $<M_{\ell\ell}<66$ GeV, and for the high mass range, $116$ GeV $<M_{\ell\ell}<1500$ GeV. I combined the two when examining this experiment. The kinematic range spans over $10^{-3}\lesssim x \lesssim 0.4$ and $2 \lesssim Q \lesssim 1000$ GeV [cf. Fig.~\ref{fig:L2_xQ2}], allowing for both low and higher $Q$ to be studied. An interesting feature exhibited in Fig.~\ref{fig:ATLASDY} is the sensitivity to $s$ and $g$ in all three PDF sets. This experiment generally favors higher magnitudes of the light-quark and gluon PDFs according to Fig.~\ref{fig:ATLASDY}. A plausible explanation for this is a missing higher-order correction, thus favoring higher theory cross sections. The larger sensitivity at large $x$ and smaller sensitivity at low $x$ reveal that the experiment sensitivities are influenced more by the high mass range rather than the low mass range. The theoretical calculations were performed at NNLO with the QCD scale set to $m_{\ell\bar{\ell}}$, but a few percent correction to the rate may be missing from the N3LO reactions that are not included. If the N3LO contributions were to be added, the sensitivities in this region could be decreased. It's worth noting that the high mass region includes NLO EW corrections, and the N3LO corrections can be either positive or negative depending on the kinematic ranges and the PDF behavior, which needs to be verified by future calculations \cite{nandakumaran2021unfolding,teney2023id}.  It is also worth noting that $L_2$ sensitivities for all three PDF sets prefer $s$ to be smaller for all $x$, and prefer a smaller gluon at high $x$. The latter feature can be explained by the conservation of momentum sum rules. All PDFs prefer to increase within the experimental kinematic region of $x\sim 10^{-3}$, and to compensate for this increase, the gluon PDF decreases in the less sensitive region of $x>0.1$. 

\subsubsection{ATLAS inclusive jet production 2.76 TeV}
\begin{figure*}
\centering
    \includegraphics[width=1.0\columnwidth]{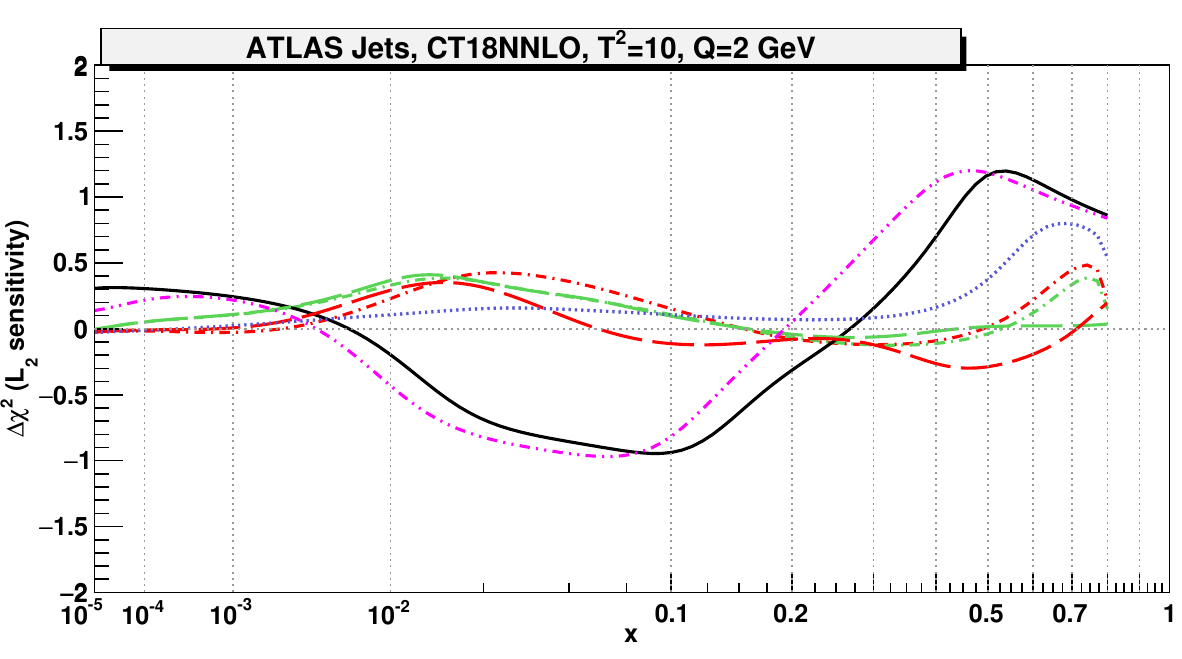}
    \def\big{\includegraphics[width=1.0\columnwidth]{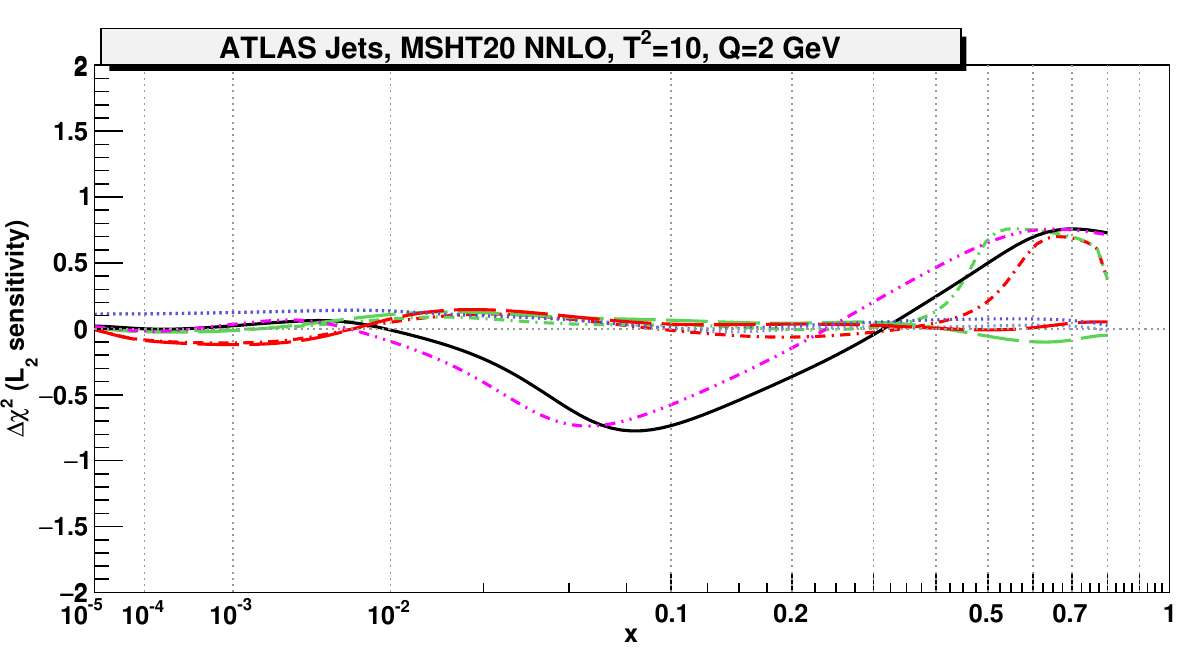}}
    \def\little{\includegraphics[width=0.15\columnwidth]{figs/L2_study/ssbarlegend.pdf}}
    \stackinset{l}{30pt}{t}{15pt}{\little}{\big}
    \caption{$L_2$ sensitivity for ATLAS jet production \cite{ATLAS:2013pbc} using the CT18, CT18As, and MSHT20 NNLO PDF sets with $T^2=10$.}
    \label{fig:ATLASjets}
\end{figure*}

The ATLAS group also measured the inclusive jet differential cross section at $\sqrt{s}=2.76$ TeV at low luminosity \cite{ATLAS:2013pbc}. The measurements cover a kinematic range of $3\times 10^{-4}\lesssim x \lesssim 0.4$ and $12 \lesssim Q \lesssim 300$ GeV [see Fig.~\ref{fig:L2_xQ2}]. In Fig.~\ref{fig:ATLASjets}, quark PDF sensitivities are minimal, while the gluon exhibits a notably negative $S^H_{f,L_2}$ for $10^{-2}\lesssim x\lesssim0.2$ and a positive $S^H_{f,L_2}$ for $x\gtrsim0.5$ in both CT18 and MSHT20. A stronger sensitivity to the gluon aligns with expectations, given the previous discussions of jet production experiments included in this study.

\subsubsection{ATLAS direct photon ratio of cross sections}

ATLAS measured the ratio of the cross sections for inclusive isolated-photon production in pp collisions at center-of-mass energies of 13 and 8 TeV. Datasets have been implemented into xFitter for a variety of ranges of $p_T$ of the photon and include the ratios $R_{\gamma}=\sigma_{\gamma}^{13 TeV}/\sigma_{\gamma}^{8 TeV}$ and $R_{\gamma}/R_Z$. In this study, I use the datasets with ranges of $125<Q<650$ GeV and $125<Q<1500$ GeV for $R_{\gamma}$. These $R_{\gamma}$ measurements were taken using the $pp\rightarrow \gamma + X$ channel \cite{ATLAS:2019drj}. The sensitivity plots for the ATLAS direct photon dataset are unremarkable. For all PDF flavors in all three PDF sets at $Q=100$ GeV, the sensitivities are flat and are centered around zero with a maximum magnitude of less than 0.5.

\section{Conclusion}
\label{conclusion}

A recent study \cite{Jing:2023isu} extensively examined sensitivities of experimental data to PDFs within the ATLAS21, CT18, CT18As, and MSHT20 PDF fits. In this article, I applied the method of Ref. \cite{Jing:2023isu} to obtain the sensitivities for data sets included in the xFitter framework, rather than in the CT18 and MSHT native fitting codes. By using the same error PDFs as in Ref. \cite{Jing:2023isu}, I compared the $L_2$ sensitivities obtained in several fitting frameworks. In my study, I explored new data sets not included in some of these PDF sets, namely, the HERA and ZEUS combined charm and beauty production, LHCb 7 TeV charm and beauty production, CMS 7 TeV W+c production, ATLAS direct photon production, and CMS 13 TeV inclusive jet production. To validate my approach using xFitter, I also compared $L_2$ sensitivities for overlapping experiments in xFitter, CT18, and MSHT20 fits with computed plots from previous works (Ref. \cite{Jing:2023isu}), while accounting for subtle differences arising from the use of different heavy-quark flavor schemes.

In my study, I have found that HERA I+II inclusive DIS is notably sensitive to $u/\bar{u}$ and gluon PDFs for all three PDF sets. H1 and ZEUS combined charm and beauty are sensitive to the quark PDFs at low $x$ as well as the gluon PDF for $x\lesssim0.7$. The CMS 7 TeV W+c production is insensitive to the quark or gluon PDFs except for the $s$ PDF, showing that the CT18As and MSHT20 PDF sets favor a smaller $s$ PDF (positive $S^H_{f,L_2}$), and CT18 favors a larger $s$ PDF (negative $S^H_{f,L_2}$). The pulls of W+c production on $s$ and $\bar{s}$ PDFs are about the same when the strangeness PDFs are allowed to be asymmetric in the CT18As and MSHT fits. For the LHCb $c$ and $b$ production experiment I imposed a cut $p_T^{meson}\geq2$ GeV to avoid the low {$x$,$Q$} region where peruturbative QCD is unstable. When these cuts are imposed, I find that the LHCb $c$ and $b$ production prefers about the same gluon for CT18 and a larger gluon for MSHT at $x<10^{-4}$. The CMS 13 TeV jet production is sensitive to the gluon for $10^{-2}<x<0.1$ for all PDF sets. The compatibility of the CMS jet data sets at 13 TeV with the 7 and 8 TeV was examined. The comparison of the 13 TeV sensitivities with the 7 and 8 TeV sensitivities reveals that the 13 TeV data is more compatible with the 8 TeV data than with the 7 TeV one. Similarly, the ZEUS jet production, H1 jet production, and ATLAS jet production experiments are sensitive to the gluon PDF for all PDF sets examined but overall have low sensitivities compared to the previously discussed experiments. The ATLAS DY pair production revealed that it favors larger quark PDFs, a larger gluon at low $x$, and smaller gluon at high $x$ for all PDF sets, and a smaller $s$ PDF for all $x$ values in CT18As, but similarly to the aforementioned three experiments, it has low sensitivities. The ATLAS direct photon production experiment is found not to be sensitive to any PDF in particular. Although some of the examined experiments are not included in the PDF sets, and someexperiments use a nonidentical factorization scheme compared to the fits, nevertheless, the agreement with the experiments is acceptable and can be improved by including them in the global fits.

xFitter is an open-source general-purpose fitting program that can find the best-fit PDFs using multiple data sets, as well as generate the error PDF sets. As a part of the output during the fitting process, xFitter returns the theoretical cross sections and $\chi_E^2$ values. xFitter demonstrates remarkable versatility, adeptly executing an array of user-defined tasks, such as computing the $L_2$ sensitivities for a plethora of experimental datasets using a wide variety of PDF sets. It should be noted that, while xFitter offers valuable capabilities, it does not supplant other methods for calculating $L_2$ sensitivities. The CTEQ code, for instance, encompasses additional features not incorporated into xFitter. By analyzing the $L_2$ sensitivities independently of the global analysis groups, it is possible to learn about the potential of the data sets in xFitter to constrain the PDF sets in the global QCD fits, such as CTEQ-TEA or MSHT.

\section*{Acknowledgment}
This work is supported in part by the U.S. Department of Energy under Grant No.~DE-SC0010129. I am grateful to P. Nadolsky for discussions on the analysis of each experiment as well as comments on the project.

%%%

\bibliographystyle{utphys}
\bibliography{biblio.bib}

\end{document}